\pdfoutput=1
\documentclass[10pt,twocolumn,compsoc]{IEEEtran}

\usepackage{alltt}
\ifCLASSOPTIONcompsoc
  \usepackage[nocompress]{cite}
\else
  \usepackage{cite}
\fi
\usepackage{cite}
\usepackage{amsmath}
\usepackage{hyperref}
 \usepackage{algorithm}
  \usepackage{algorithmic}
\usepackage{multirow}
\usepackage{enumitem}
\usepackage{bigstrut}
\usepackage{balance}
\usepackage{wrapfig}
 \usepackage[flushleft]{threeparttable}
 
 \usepackage{colortbl}
 
\usepackage[dvipsnames]{xcolor}
 \usepackage{dblfloatfix}
 
\usepackage[usestackEOL]{stackengine}
\usepackage{xcolor, graphicx}

\newcommand{\IT}[1]{{\bf%
DODGE(\ifx*#1$\mathcal{E}$\else#1\fi)}}
\newcommand{\tion}[1]{\S\ref{sect:#1}}

\newcommand{\fig}[1]{Figure~\ref{fig:#1}}

\newcommand{\tbl}[1]{Table~\ref{tbl:#1}}
\newcommand{\bi}{\begin{itemize}[leftmargin=0.4cm]}
	\newcommand{\ei}{\end{itemize}}
\newcommand{\be}{\begin{enumerate}[leftmargin=0.4cm]}
	\newcommand{\ee}{\end{enumerate}}

\newcommand{\citeresp}[1]{}
\newcommand{\respto}[1]{}

\newcommand{\revised}{\textcolor{black}}

\newcommand{\BLUE}{\color{black}}
\newcommand{\BLACK}{\color{black}}

\begin{document}
 
\title{Simpler Hyperparameter Optimization for\\ Software Analytics: Why, How, When?} 

\author{Amritanshu Agrawal, Xueqi Yang, Rishabh~Agrawal,
Rahul Yedida, Xipeng Shen,
        Tim~Menzies 
\IEEEcompsocitemizethanks{\IEEEcompsocthanksitem A. Agrawal works at Wayfair  email: aagrawa8@ncsu.edu.
The other authors are at the Department of Computer Science, North Carolina State University, Raleigh, USA.
\{xyang37, ragrawa3,
xshen5\}@ncsu.edu,  yrahul3910@gmail.com, timm@ieee.org}}

\markboth{IEEE Transactions on Software Engineering}%
{Agrawal \MarkLowerCase{\textit{et al.}}: Simpler Software Analytics: Why, How, and When?}


\IEEEtitleabstractindextext{
\begin{abstract}
How can we make software analytics simpler and faster?
One method is to match the complexity of  analysis to the intrinsic complexity of the data being explored. For example,
hyperparameter optimizers
find  the  control settings for  data miners that improve  the predictions generated via software analytics.
\respto{1.4} \revised{Sometimes,  very fast hyperparameter optimization can  be achieved by
 ``DODGE-ing''; i.e. simply  steering way from   settings that
 lead to similar conclusions.  }
 \respto{3.2c} \revised{
But when is it wise to use that simple approach and when must we use more complex (and much slower)   optimizers?}
To answer this, we applied hyperparameter optimization to 120 SE data sets that explored bad smell detection, predicting Github issue close time,  bug report analysis, defect prediction, and dozens of other non-SE problems. 
We find that the simple DODGE works best for data sets with low ``intrinsic dimensionality''  ($\mu_D\approx 3$) and very poorly for higher-dimensional data ($\mu_D > 8$).
Nearly all the SE data  seen here was  intrinsically low-dimensional, indicating that DODGE is applicable for many SE analytics tasks.

 
\end{abstract}

}

\maketitle





\section{Introduction}

\IEEEPARstart{I}{ndustrial} practitioners 
(and researchers) use data mining and software analytics for many tasks~\cite{wan18,export:208800,theisen15,menzies2016perspectives,gall2014software,bird2015art}
such as
learning
how long it will take to integrate  new code~\cite{czer11};
where bugs are most likely~\cite{ostrand2004bugs,me07e}; or how long it will take to develop this code~\cite{me11a,koc11b}.
Even simple design decisions
such as the color of a link are chosen by analytics~\cite{linares2014mining}.
In industrial settings, software
analytics can be remarkably
 cost-effective~\cite{misirli2011ai,kim2015remi}.
 Also,    such analytics
  perform competitively
   with seemingly more rigorous approaches like static code analysis~\cite{rahman2014comparing}.  

\begin{wrapfigure}{r}{1.8in}

 \includegraphics[width=1.8in]{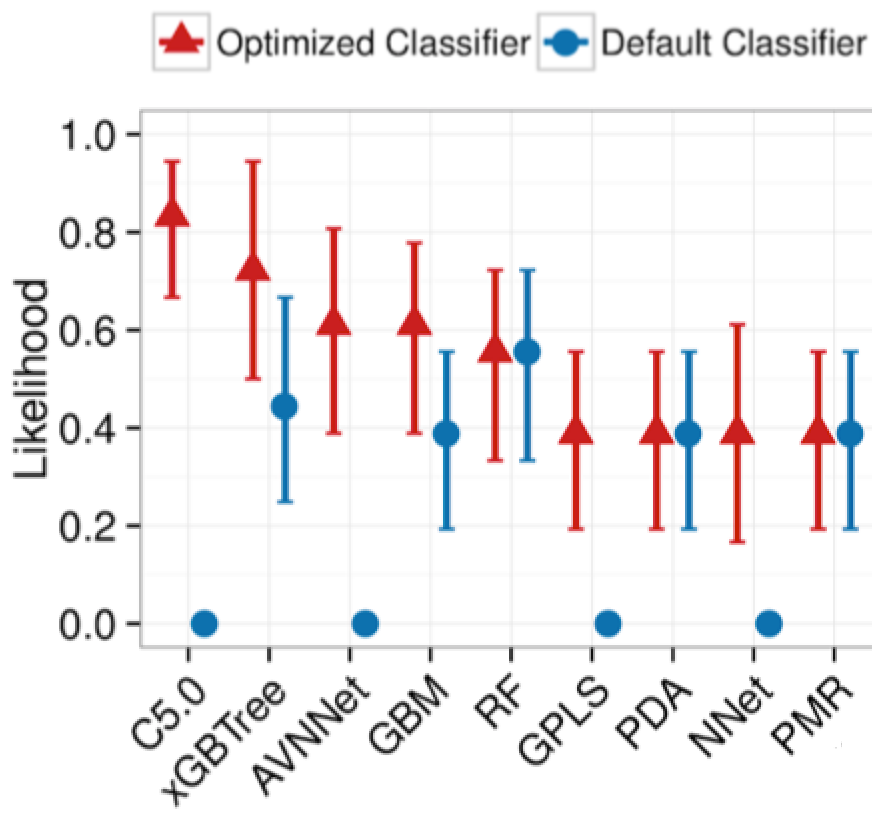}

\caption{Optimization   from~\cite{tantithamthavorn2016automated}.
Y-axis shows   likelihood  a learner   performs best. 
Blue/red shows  results  before/after optimization.
Before tuning, C5.0 (bottom-left) seems unpromising. But afterwards, it   is best.}\label{fig:tanit}
\end{wrapfigure}

One of the black arts of software analytics is how to set the ``magic parameters'' that control a learner (e.g. how many clusters should we hunt for?).
 Hyperparameter optimizers are automatic tools that find ``good''  settings for  data miners  Here by ``good'' we mean that those
settings can  greatly improve prediction accuracy for software analytics~\cite{czer11,ostrand2004bugs,me07e,me11a,koc11b,linares2014mining,rahman2014comparing,liu2010evolutionary,sarro2012further,zhong2004,treude2018per,oliveira2010ga}.
For example, Tantithamthavorn et al.~\cite{tantithamthavorn2016automated} showed that such optimizers can convert 
very bad learners into outstandingly good ones (see the gains observed in  \fig{tanit}).   But hyperparameter optimization can be 
very  slow. \tbl{options} shows
some of the   hyperparameter options seen in recent SE papers~\cite{agrawal2019dodge}. 
Assuming that the numerics of that table divide into ten bins, then \tbl{options} lists billions of   possibilities. 

Recently, we achieved  success using a surprisingly simple hyperparameter optimizer called DODGE~\cite{agrawal2019dodge} that
``dodges'' away from  (i)~options  tried before and which
have  (ii)~resulted in
similar performance scores. More about DODGE in \S\ref{sect:relax}.
DODGE ran orders of magnitude faster than prior methods since its search terminated after 30 evaluations (while  other methods used thousands to millions of options). Also, its results were as good, or better, than prior state-of-the-art results.

 A  deficiency in those prior results is that it only
gave examples where DODGE {\em worked}, but not when it {\em failed}.
No optimizer works best on all data~\cite{Wolpert97}. 
Accordingly, 
\respto{3.2b} \revised{the goal of this paper is to determine under what conditions we should
use DODGE or when should we use more complex (and much slower) methods. }
 
The next section 
motivate  {\em \underline{why}} it is so important to  seek simpler software analytics.
 \begin{table}[!t]
\caption{Hyperparameter  options
seen in recent SE papers~\cite{ghotra2015revisiting,fu2016tuning,agrawal2018better,agrawal2018wrong} and in  the documentation of a widely-used data mining library
(Scikit-learn~\cite{scikit-learn}).
}\label{tbl:options}
\scriptsize
 
\begin{tabular}{|p{.95\linewidth}|}\hline
\textbf{Learners:}

\noindent
\bi
\item DecisionTreeClassifier(criterion=b, splitter=c, min\_samples\_split=a)
    \bi
   \item a, b, c= randuniform(0.0,1.0), randchoice([`gini',`entropy']), \newline randchoice([`best',`random'])
   \ei
\item RandomForestClassifier(n\_estimators=a,criterion=b,  min\_samples\_split=c)
    \bi
   \item a,b,c = randint(50, 150), randchoice(['gini', 'entropy']),\newline randuniform(0.0, 1.0) 
   \ei
\item LogisticRegression(penalty=a, tol=b, C=float(c))
    \bi
    \item a,b,c=randchoice([`l1',`l2']), randuniform(0.0,0.1), randint(1,500)
    \ei 
\item MultinomialNB(alpha=a) = randuniform(0.0,0.1)
\item KNeighborsClassifier(n\_neighbors=a, weights=b, p=d, metric=c)
    \bi
    \item a, b,c  = randint(2, 25), randchoice([`uniform', `distance']), \newline randchoice([`minkowski',`chebyshev'])
    \item if c=='minkowski': d= randint(1,15)  else:  d=2
    \ei

\ei
\\\hline

{\bf Pre-processors for   defect prediction, Issue lifetime, Bad   Smells and Non-SE:} 

\bi 
\item StandardScaler, MinMaxScaler, MaxAbsScaler
\item RobustScaler(quantile\_range=(a, b)) =  randint(0,50), randint(51,100) 
\item KernelCenterer
\item QuantileTransform(n\_quantiles=a,  output\_distribution=c, subsample=b)
    \bi
    \item a, b = randint(100, 1000), randint(1000, 1e5)
    \item c = randchoice([`normal',`uniform'])
    \ei
\item Normalizer(norm=a) = randchoice([`l1', `l2',`max'])
  
\item Binarizer(threshold=a) =  randuniform(0,100)
   
    \item SMOTE(n\_neighbors=, n\_synthetics=b,  Minkowski\_exponent=c)
    \bi
    \item a,b = randit(1,20),randchoice(50,100,200,400)
    \item c = randuniform(0.1,5)
    \ei
\ei

 \\\hline
{\bf Pre-Processors for Text mining:}  
 
\bi
\item CountVectorizer(max\_df=a, min\_df=b) = randint(100, 1000), randint(1, 10)
\item TfidfVectorizer(max\_df=a, min\_df=b, norm=c)
    \bi
    \item a, b,c = randint(100, 1000), randint(1, 10), randchoice([`l1', `l2', None])
    \ei
\item HashingVectorizer(n\_features=a, norm=b)
    \bi
    \item a = randchoice([1000, 2000, 4000, 6000, 8000, 10000])
    \item b = randchoice([`l1', `l2', None])
    \ei
\item LatentDirichletAllocation(n\_components=a, doc\_topic\_prior=b,\newline topic\_word\_prior=c,
                                   learning\_decay=d,
                                   learning\_offset=e,batch\_size=f)
    \bi
    \item a, b, c = randint(10, 50), randuniform(0, 1),  randuniform(0, 1)
    \item d, e  = randuniform(0.51, 1.0), randuniform(1, 50), 
    \item f = randchoice([150,180,210,250,300])
    \ei
\ei

 \\\hline
 \end{tabular}

\end{table}
After that, we look into
 {\em \underline{how}} to simplify the computational cost of hyperparameter optimization in software analytics.
 Algorithms for hyperparameter optimization are discussed. 
  These are  
 tested on 120 data sets from five domains:
 four from SE and one ``miscellaneous set'' of non-SE data
 taken from the UCI machine learning repository~\cite{Dua:2019}.
Finally, we check
{\em \underline{when}}   simplification is possible.
Our 120  data sets can be characterized by their ``intrinsic dimensionality'' which measures $D$, the number of underlying
dimensions in a data set. DODGE performs best for data sets with low dimensionality ($\mu_D\approx 3$) and very poorly for higher-dimensional data ($\mu_D > 8$).
Nearly all the SE data explored here were low-dimensional, indicating that DODGE's simple analysis is applicable for many SE analytics tasks.

\subsection{Connection to Prior Work}
This paper is a significant extension to prior work, in two ways.
Here,  we explore more data from more SE domains than prior studies.
Previously~\cite{agrawal2019dodge},   
DODGE was assessed using 16 data sets from just two domains:
\be
\item
 10 SE defect prediction data sets;
 \item
 6 SE issue tracking data sets.
 \ee
 This study repeats that analysis while also studying
 \be
\item[3)]
 63    SE data sets exploring Github issue close time;
 \item[4)]
 4    SE data sets exploring bad smell detection.
 \item[5)]
37  non-SE problems
 from the UCI repository.
 \ee
But more importantly, here we show that intrinsic dimensionality can predict when DODGE will {\em not work}. This is a useful result since
intrinsic dimensionality can be applied to data before starting an analysis. That is, now we can determine when to use DODGE, or some other method,
before analysts waste any time applying the wrong optimizer.

\subsection{Reproduction Package}

To encourage reproduction of this work, all our code and scripts  are  available on-line at \url{http://tiny.cc/dodge2020}.

\revised{\subsection{Background Notes}
\respto{2.2} \respto{3.5}
This paper explores  a   wide range of technologies. Hence, before we begin, it is   appropriate to  offer some
introductory notes on that material.}

\revised{In this paper,
{\em data} are tables with {\em rows} and {\em columns}.
Columns are also known as 
{\em features, attributes}, or 
{\em variables}.}

\revised{Rows contain multiple $X,Y$ features where $X$ are the  {\em independent variables} (that can be observed, and sometimes controlled) while $Y$ are the {\em dependent} variables (e.g. number of defects). When $Y$ is absent,  then {\em unsupervised}  learners seek mappings
between the $X$ values. For example,
{\em clustering} algorithms find groupings of similar rows (i.e. rows with similar $X$ values).}

\revised{Usually most rows have values for most $X$ values. But with {\em text mining}, the opposite is true. In principle, text miners have one column for each work in text's language. Since not all documents use all words, these means that the rows of a text mining data set are often ``sparse''; i.e. has mostly missing values.}

\revised{When $Y$ is present and there is only one of them (i.e. $|Y|=1$) then {\em supervised} learners seek mappings from the $X$ features to the $Y$ values. For example,
{\em logistic regression} tries to fit the $X,Y$ mapping to a particular equation. }

\revised{When there are many $Y$ values (i.e. $|Y|>1$), then another array $W$ stores a  set of 
{\em weights} indicating what we want to  minimize or maximize  (e.g. we
would seek to minimize $Y_i$ when \mbox{$W_i<0$}).  In this case,
{\em multi-objective optimizers} seek $X$ values that most minimize or maximize their associated $Y$ values. So:
\bi
\item
{\em Clustering} algorithms find groups of rows;
\item
and {\em Classifiers}
(and {\em regression algorithms}) find how those groups relate to the target $Y$ variables; 
\item
\BLUE
and {\em Optimizers}
are tools that suggest ``better'' settings
for the  $X$ values (and, here, ``better'' means settings that improve the expected value
of the  $Y$ values).\BLACK
\ei
Apart from $W,X,Y$,  we add $Z$, the {\em hyperparameter} settings that control how learners performs regression or clustering.
For example, a KNeighbors algorithm needs
to know how many nearby rows to use for its classification (in which case, that $k\in Z$).
Usually the $Z$ values are shared across all rows (exception: some optimizers first cluster the data and use different $Z$ 
settings for different clusters).}

\revised{Two important detail not discussed above are  {\em feature engineering} 
and  how to select {\em performance metrics}.
Feature engineering includes all the pre-processing algorithms listed in 
Table~\ref{tbl:options}. These algorithms
are used to ``massage'' data prior to clustering or classification or optimization.
For example,   the LDA pre-processor shown in Table~\ref{tbl:options} is a text mining pre-processor that finds {\em topics}; i.e. words that often occur together within the same paragraph. Topics usually occur at  exponentially decreasing frequency; i.e. , a dozen or so topics might cover most of the document space. Later in this paper,
we will   (a)~replace sparse raw text mining data with a non-sparse ``topics matrix'' comprising one column per topics and rows showing how much each document matches each topic; then (b)~run a simple learner over this non-sparse matrix.}

\respto{3.2} \revised{As to {\em performance metrics}, these are discussed in detail in \S\ref{sect:pm}. 
Though different performance metrics are adjusted depending on what domain is explored as described below. 
For example,
{\em Popt(20)} is a performance metric that is maximal when
a defect detector finds the  fewest lines of code with the most defects. This is useful for defect prediction  (and not other domains) since the business justification for defect prediction is ``do not ask us inspect too much code''. }

\section{Why? On the Value of Simpler  Analytics}
 
To  motivate this work, we must first
 explain why it is so important to seek simpler software analytics.
 
Fisher et al.~\cite{fisher12} characterizes software analytics as a workflow that distills large quantities of low-value data down to smaller sets of higher-value data.  
Hyperparameter optimization  improves the predictions generated by software analytics,
but it also increases the computational cost of software analytics.
Fisher et al.~\cite{fisher12} warn against any such increase. They say:

  \begin{quote}
{\em Further advances in advanced software analytics will be stunted unless
we can tame their associated CPU costs}.
  \end{quote}
They note that
due to the complexities and computational cost of SE analytics,
``the
luxuries of interactivity, direct
manipulation, and fast system
response are gone''~\cite{fisher12}. They
characterize modern
cloud-based analytics as a throwback to the 1960s--
batch processing
mainframes where jobs are submitted and then analysts wait and wait
for results  with ``little insight
into what's really going on behind
the scenes, how long it will take, or
how much it's going to cost''~\cite{fisher12}.

Fisher et al. document the issues seen by 16 industrial data scientists,
one of whom remarks ``Fast iteration is
key, but incompatible with the way jobs are submitted and processed
in the cloud. It's frustrating to wait for hours, only to realize you need a 
slight tweak to your feature set.''.

\begin{figure}[!b]
\footnotesize
\begin{alltt}
 1. {\bf FOR} \(D\)  data sets {\bf DO} # for each data set do
 2.   {\bf FOR} \(R=5\) times {\bf DO}   
 3.     Randomly divide  data to \(B=5\) bins;
 4      {\bf FOR} \(i=1..B\) {\bf DO}
 5         test  = bin{[}i{]}
 6         train = data - test
 7.        {\bf FOR} \(F\) options from \tbl{options} {\bf DO}
 8.            model = \(F\)(data)  
 9.            print report(apply(model,test))
\end{alltt}

\caption{Software analytics evaluation.}\label{fig:eval}
\end{figure}

To understand the CPU problem
consider the standard validation loop for a data miner shown in \fig{eval}.
Note the problem with this loop---it must call a data miner (at line 8)  $D*R*B*F$ times. This is a problem since:
\bi
\item
$D$ is an ever increasing number. 10 years ago, a paper on software
analytics could be published if it used $D<10$ data sets. Now, as shown in this paper, it is common to see papers with $D>10^2$ data sets. In the future, as more data is extracted from open source projects (e.g. those found in
Github), we expect that using $D>10^3$  data sets will be common.
\item
It is usual for $R*B>20$ since, 
for statistical validity, it is common to repeat this loop more
than 20 times.  
\item
 $F$ comes from \tbl{options}.
Assuming  that  the  numerics  of  \tbl{options} are  divided into  ten  bins,  then  \mbox{$F>10^9$}.   Since this number is too large to be explored, it is common practice to use ``engineering judgement'' (a.k.a.  guessing) to reduce $F$ to
$10^6$ or $10^3$.
\ei
Even after imposing engineering judgement, the inner loop of \fig{eval} must call a learner millions to  billions of times.
This is troubling since while
some data miners  are very fast (e.g. Naive Bayes), some are not (e.g. deep learning).
Worse still, several ``local learning'' results~\cite{me12d} report that software analytics results are specific to the data set
being processed---which means that analysts may need to rerun the above loop  anytime new data comes to hand.


Note that this CPU problem is not solvable by (1)~parallelization or (2)~waiting for faster CPUs.
Parallelization requires the kinds of environments that Fisher et al. discuss; i.e. environments where it is frustrating to wait for hours, only to realize you need a slight tweak to your
feature setting. 
As to waiting for faster CPUs,  
it is not clear that we can  rely on Moore's Law~\cite{moore65} to double our computational power every 18 months.
Power consumption and heat dissipation issues effectively block
further exponential increases to CPU clock frequencies~\cite{kuman03}.


\section{How?  Hyperparameter Optimization}
\label{sect:methods}

In this section, first we discuss learning algorithms. Next, we discuss methods
for learning the control settings for those learners.

\subsection{Data mining tools}\label{sect:standard}

Hyperparameter optimizers adjust the control parameters of data miners. This
section
reviews
the machine learning algorithms used in this study: 
 SVM,
 Random Forests,
 decision tree learners,
 logistic regression,
 Naive Bayes,
 and LDA.

Before doing that, it is reasonable to ask ``why did we select these tools, and not some other set?''.
This paper does not compare DODGE against all other learners and all other hyperparameter optimizers (since such a comparison would not fit into a single paper). Instead, we use baselines as found in the SE literature for bad smell detection, predicting Github issue close time, bug report analysis, and defect prediction.
    
For example, for defect prediction,  our classifiers come from a study by Ghotra et al.~\cite{ghotra2015revisiting}. They found that the  performance of dozens of data miners (applied to defect prediction) can be clustered into just a few groups. By sampling a few algorithms from each group, we can explore the range of data miners seen in defect prediction.

Clustering algorithms like EM~\cite{Dempster77maximumlikelihood} divide the data into related groups, then check the properties of each group. Another clustering method used in text mining, is Latent Dirichlet Allocation~\cite{Blei:2003} that infers ``topics'' (commonly associated words). After documents are scored according to how often they use some topic, a secondary classifier can then be used to distinguish the different topics.

Clustering algorithms like EM and LDA  might  not make use of any class variable.
Naive Bayes classifiers~\cite{Duda:2000}, on the other hand, always divide the data on the class. New examples are then classified according to which class it is most similar to.
Also, logistic regression fits the data to a particular parametric form (the logistic function).

Another learner that uses  class variables are decision tree algorithms~\cite{Quinlan1986,breiman2017classification}. 
These learners divide data on attributes whose values most separate the classes and then recurses on each division. Random Forests~\cite{breiman2001random} build a ``committee'' of multiple decision trees, using different sub-samples of the data. Conclusions then come from a voting procedure across all the  trees in the forest. \respto{3.7} \revised{Distance-based classifiers like KNN, one the other hand, classify test data by looking at the test instance's ``k' nearest neighbors.}

Standard clustering and decision tree algorithms base their analysis using the raw problem data. But what if some extra derived attribute is best at separating the classes? To address that issue, SVMs use a ``kernel'' to infer that extra dimension~\cite{Boser:1993}.  

\respto{3.3} \revised{All these algorithms have their own particular hyperparameters.
In our work, we select those hyperparameters using two methods:
\bi
\item For our state-of-the-art ``SOTA'' studies (defined in the next section),   we used
the default parameters from the SCIKIT-LEARN toolkit. This is a 
  widely-used toolkit in the software analytics domain (see~\cite{scikit-learn}, 34,726 citations in Google Scholar since 2011). 
  \item For our other``TPE'' studies
  (also defined below), we select
  hyperparameters automatically using two methods: DODGE and the HYEROPT system described in the next section.
  \ei}

\subsection{Hyperparameter Optimizers}\label{HPO}
\revised{In this paper, we will assess  DODGE against
``SOTA'' and ``TPE'':
\bi
\item  SOTA is our shorthand for the prior state-of-the-art seen in the SE literature for different domains.
\item TPE is short for tree-structured Parzen estimators  which is 
a state-of-the-art optimizer taken from the AI
literature. To the best of our knowledge, this
algorithm has not previously been applied to any SE
data sets.
\ei}
\noindent
Different SE domains use different SOTA algorithms.
For example,
for text mining SE data,   Panichella et al.~\cite{Panichella:2013} used genetic algorithm~\cite{goldberg1988genetic} (GA)  to ``evolve'' a set of randomly generated control settings for SE text miners by repeating the following procedure, across many ``generations'': (a)~mutate  a large population of alternate settings;
 (b) prune the worse performing settings;   (c)~combine pairs of the better, mutated options.
 
Two other SE SOTA hyperparameter optimization algorithms are    differential evolution~\cite{storn1997differential} and grid search.
Different evolution is used by Fu et al.~\cite{fu2016tuning} and Agrawal et al.~\cite{agrawal2018better}.
DE generates mutants by interpolating between the better-ranked settings.
These better settings are kept in a ``frontier list''. Differential evolution iterates over the frontier, checking each candidate against
a new mutant. If the new mutant is better, it replaces the frontier item, thus improving the space of examples
used for subsequent mutant interpolation.
Tantithamthavorn et al.~\cite{tantithamthavorn2016automated} used a grid search for their hyperparameter optimization study. Grid search runs nested ``for-loops'' over the range of each control option.  Fu et al.~\cite{Fu16zzz} found that for defect prediction, grid search ran 100 to 1000 times slower than DE.
 
\respto{1.1a} \revised{As to   state-of-the-art optimizers from outside the SE  
literature, 
a December 2020 Google Scholar search for ``Hyperparameter optimization'' 
 reported that two  papers by 
 Bergstra et al.~\cite{bergstra2012random,bergstra2011algorithms} have most citations (2159 citations  and 4982 citations\footnote{The nearest other work was a 2013 paper by Thornton et al. on Auto-WEKA~\cite{thornton2013auto} with 931 citations.}). Accordingly,
 when we compare DODGE against an algorithm
 not developed by the SE community, we will use
 the  TPE algorithm recommended by 
  Bergstra et al. For our experiments,
  we use the TPE implementation from Bergstra's HYPEROPT toolkit~\cite{10.5555/3042817.3042832}.}
  
 \revised{ TPE reflects over the evaluations seen to date
  in order to select the next best setting to explore. More specifically:
  \bi
  \item
 TPE takes the evaluations made so far and divides them into two groups: {\em best} and 
  {\em rest}. 
  \item
  Each group is then modelled as a Gaussian with its own mean and standard deviation.
  \item
  A stochastic method then proposes  random hyperparameter settings.
  \item Before running those options, TPE prunes the proposed settings that belong least to {\em rest}  (so most likely to belong to {\em best}).\ei }

\newlength\sqsz
\sqsz=.3cm

\newcommand\sq[1]{\protect\sqhelper{#1}}

\newcommand\sqhelper[1]{\sffamily\bfseries\fboxsep=-\fboxrule%
  \ifnum#1>7\def\sqcolor{green}\def\sqcat{~}\else%
    \ifnum#1>2\def\sqcolor{lightgray}\def\sqcat{~}\else%
      \def\sqcolor{red}\def\sqcat{~}\fi\fi%
  \fcolorbox{black}{\sqcolor}{\makebox[\sqsz]{%
  \rule[\dimexpr-.5\sqsz+.2\ht\strutbox]{0pt}{\sqsz}\stackanchor[2pt]{~}{~}}}}

\newcommand\hsq[1]{\protect\hsqhelper{#1}}

\newcommand\hsqhelper[1]{\sffamily\bfseries%
  \makebox[\sqsz]{\rule{0pt}{\dimexpr.5\sqsz+.2\ht\strutbox}#1}}

\newcommand\vsq[1]{\protect\vsqhelper{#1}}

\newcommand\vsqhelper[1]{\sffamily\bfseries%
  \rule[\dimexpr-.5\sqsz+.2\ht\strutbox]{0pt}{\sqsz}#1}

\subsubsection{Optimizing with DODGE}
\label{sect:relax}
 
DODGE is a hyperparameter optimizer proposed by  Agrawal et al.~\cite{agrawal2019dodge}.  
DODGE was designed around the following observation.
Given an ever-evolving set of tools, languages, platforms, tasks, user expectations, development population,
development practices, etc, we might expect that any prediction about an SE project will only ever be approximately accurate, i.e., within   $\epsilon$ of the true value.
Agrawal et al. reasoned that     $\epsilon$ is not a problem
 to be solved, but a resource that could be exploited, as follows:
 \begin{quote}
{\bf The RELAX heuristic:}
{\em Ignore anything less than $\epsilon$.}
\end{quote}
DODGE applies this RELAX heuristic to do hyperparameter optimization. To
illustrate this process, consider the following example:
\bi
\item
Suppose we are exploring the hyperparameter space of \tbl{options};
\item
Suppose further we are scoring each hyperparameter setting by applying it to a learner, then recording
the   performance goals of  recall and false alarm  seen after applying
those settings to a learner.
\ei
 
  \begin{wrapfigure}{r}{1.5in}
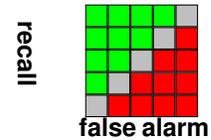

 \begin{center}
\raisebox{.5\sqsz}{\rotatebox{-90}{\sffamily\bfseries\makebox[4\sqsz]{recall}}}~
\setstackgap{S}{0pt}
\Shortunderstack{\vsq{ }\\ \vsq{ }\\ \vsq{ }\\ \vsq{ }\\ \vsq{ }}~~
\stackunder[9pt]{%
  \Shortunderstack{
    \sq{16}\sq{12}\sq{8}\sq{8}\sq{4}\\
    \sq{12}\sq{8}\sq{8}\sq{4}\sq{1}\\
    \sq{8}\sq{8}\sq{4}\sq{1}\sq{1}\\
    \sq{8}\sq{4}\sq{1}\sq{1}\sq{1}\\
    \sq{4}\sq{1}\sq{1}\sq{1}\sq{1}\\
    \hsq{false}\hsq{\vspace{10mm}\hspace{12mm}alarm}\hsq{}\hsq{~}
  }
}{~}
\end{center}
\caption{25 cells (if $\epsilon=0.2$).}\label{fig:cells}
\end{wrapfigure}
Given performance goals with the range $0\le g \le 1$, 
$\epsilon$~divides the
performance output space  into $(1/\epsilon)^g$ cells.
For example, consider
the $g=2$ goals of
recall and false alarm. These  have minimum and maximum values
 of zero and one. Hence, if
  $\epsilon=0.2$, then
 these scores divide into five regions   (at 0.2, 0.4, 0.6, 0.8).
 As shown in 
\fig{cells},  these divided scores separate
  a two-dimensional plot of  recall vs false alarm scores  into $(1/0.2)^2=25$ cells. In those
cells, green denotes good performance (high recall, low false alarm) and red denotes cells with relatively worse performance.

When billions of inputs (in \tbl{options}) are mapped into   the 25 cells of \fig{cells}, then   many  inputs  are {\em redundant}, i.e.,    lead to the   same outputs.  
The faster we can ``dodge''   redundant options, the faster we can move on to  explore the other $(1/\epsilon)^g$   possible outputs. 

To implement ``dodging'',
 DODGE models \tbl{options}
 as a tree  where all nodes have initial weights $w=0$.
Next,  $N_1$ times, DODGE selects branches at random.
We evaluate the options in a branch and if the resulting scores are within   $\epsilon$ of any previous scores, then DODGE deprecates those options via \mbox{$w=w-1$}, else $w=w+1$. 

After that,
DODGE freezes the selected branches found so far.  
$N_2$ times, DODGE then makes random selection to restrict
 any numeric ranges. 
When a  range is initially  evaluated, a random number \mbox{$r=random(\mathit{lo}, \mathit{hi})$} is selected and its weight $w(r)$ is set to zero.
Subsequently, this 
weight is adjusted (as described above).
When  a new value  is required
(i.e., when the branch is evaluated again) then  if the  best, worst weights seen so far (in this range) are
 $x,y$ (respectively) then we reset $\mathit{lo},\mathit{hi}$
 to:
 
{\footnotesize
 \[
 \begin{array}{c@{~}c@{~}c@{~}c@{~}c@{~}c}
\mathit{IF} &  x \le y &\mathit{THEN} & \mathit{lo},\mathit{hi} = x,(x+y)/2
    &\mathit{ELSE} & \mathit{lo},\mathit{hi} = (x+y)/2,x
\end{array}
 \]}
\respto{3.2a} \revised{When $\epsilon$ is large,   a few samples should suffice to find good results.  Hence,
 Agrawal et al.~\cite{agrawal2019dodge} recommends
  $\epsilon=0.2$   and $N_1 = N_2 = 15$. }
  
  DODGE can be recommended for two reasons. Firstly, for SE problems, DODGE's  optimizations   are 
  better than the prior state-of-the-art (evidence: see ~\cite{agrawal2019dodge}, and the rest of this paper). 
  
  Secondly,
  DODGE
  achieves those results very quickly.
Based on the default parameters suggested by     Goldberg~\cite{goldberg1988genetic}, Storn~\cite{storn1997differential}, and using some empirical results from Fu et al.~\cite{fu2016tuning},
we can compute the number of times a hyperparameter optimizer would have to call a data miner.
 If we are analyzing ten data sets 25 times\footnote{Why 25? In a  5x5 cross-val experiment, the data set order is randomized five times. Each time, the data is divided into five bins. Then, for each bin,
 that bin becomes a test set of a model learned from the other  bins.},
 then hyperparameter optimization with grid search or genetic algorithms or differential evolution would need to call a data miner thousands to millions of times (respectively).


\subsection{Data Used to Assess DODGE}\label{sect:data}

As stated in the introduction,
previously, DODGE   
was assessed using 16 data sets from two domains~\cite{agrawal2019dodge}:
\be
\item
 10 SE defect prediction data sets;
 \item
 6 SE issue tracking data sets.
 \ee
 This study repeats that analysis while also studying
 \be
\item[3)]
 63 SE data sets exploring Github issue close time;
 \item[4)]
 4    SE data sets exploring bad smell detection.
 \item[5)]
 Finally, we also explore 37 miscellaneous non-SE problems
 from the UCI repository\footnote{For decades, this UCI repository has been the standard source
 of data used by the machine learning community
 in their research papers~\cite{frank2010uci}.}.
 \ee
We will find that DODGE works very well
for SE case studies and very badly for  non-SE case studies. 
Later in this paper, we precisely characterize the kinds of data for which DODGE is not recommended.

The rest of this section describes the data from these five different categories.



\subsubsection{Defect Prediction}\label{sect:dp}

Software developers are smart, but sometimes make mistakes. Hence, it is essential to test the software before the deployment ~\cite{orso2014software,barr2015oracle,yoo2012regression, myers2011art}. 
Software bugs are not evenly distributed across the project~\cite{hamill2009common,koru2009investigation, ostrand2004bugs,misirli2011ai}.  Hence, 
a useful way to perform software testing is to allocate most assessment budgets to the more defect-prone parts in software projects.
Data miners can learn a predictor for defect proneness using, e.g., the static code metrics of \tbl{static_metrics}.



\begin{table}[!t]
\caption{Static  code metrics   for defect prediction.
	    For   details, see~\cite{krishna17a}.}\label{tbl:static_metrics}
	\renewcommand{\baselinestretch}{0.7}\begin{center}
	 \footnotesize
			\begin{tabular}{c|l}
				amc & average method complexity \\\hline
				avg\, cc & average McCabe \\\hline
				ca & afferent couplings \\\hline
				cam & cohesion amongst classes \\\hline
				cbm & coupling between methods \\\hline
				cbo & coupling between objects \\\hline
				ce & efferent couplings \\\hline
				dam & data access\\\hline
				dit & depth of inheritance tree\\\hline
				ic & inheritance coupling\\\hline
				lcom (lcom3) & 2 measures of lack of cohesion in methods \\\hline
				loc & lines of code \\\hline
				max\, cc & maximum McCabe\\\hline
				mfa & functional abstraction\\\hline
				moa &  aggregation\\\hline
				noc &  number of children\\\hline
				npm & number of public methods\\\hline
				rfc & response for a class\\\hline
				wmc & weighted methods per class\\\hline
			 
				defects & Boolean: where defects found in bug-tracking\\
			\end{tabular}
	 
	\end{center}
	
\end{table}

\begin{table}[!t]
\centering
\caption{Defect prediction data from http://tiny.cc/seacraft. Uses metrics from \tbl{static_metrics}. } 
\label{tbl:versions}
\scriptsize
\begin{tabular}{cllll}
\hline
\rowcolor{lightgray} \multicolumn{1}{c|}{}& \multicolumn{2}{c|}{Training Data}& \multicolumn{2}{c}{Testing Data}\\ \cline{2-5} 
\rowcolor{lightgray}\multicolumn{1}{c|}{\multirow{-2}{*}{Project}} & \multicolumn{1}{c|}{Versions} & \multicolumn{1}{c|}{\% of Defects} & \multicolumn{1}{c|}{Versions} & \multicolumn{1}{l}{\% of Defects} \\ 
\hline

\multicolumn{1}{l}{Poi}& \multicolumn{1}{l}{1.5, 2.0, 2.5}& \multicolumn{1}{l}{426/936 = 46\%}& \multicolumn{1}{c}{3.0}& \multicolumn{1}{l}{281/442 = 64\%} \\ \hline
\multicolumn{1}{l}{Lucene}& \multicolumn{1}{l}{2.0, 2.2}& \multicolumn{1}{l}{235/442 = 53\%}& \multicolumn{1}{c}{2.4}& \multicolumn{1}{l}{ 203/340 = 60\%} \\ \hline
\multicolumn{1}{l}{Camel}& \multicolumn{1}{l}{1.0, 1.2, 1.4}& \multicolumn{1}{l}{374/1819 = 21\%}& \multicolumn{1}{c}{1.6}& \multicolumn{1}{l}{ 188/965 = 19\%
} \\ \hline
\multicolumn{1}{l}{Log4j}& \multicolumn{1}{l}{1.0, 1.1}& \multicolumn{1}{l}{71/244 = 29\%}& \multicolumn{1}{c}{1.2}& \multicolumn{1}{l}{189/205 = 92\%} \\ \hline
\multicolumn{1}{l}{Xerces}& \multicolumn{1}{l}{1.2, 1.3}& \multicolumn{1}{l}{140/893 = 16\%}& \multicolumn{1}{c}{1.4}& \multicolumn{1}{l}{437/588 = 74\%} \\ \hline
\multicolumn{1}{l}{Velocity}& \multicolumn{1}{l}{1.4, 1.5}& \multicolumn{1}{l}{289/410 = 70\%}& \multicolumn{1}{c}{1.6}& \multicolumn{1}{l}{78/229 = 34\%} \\ \hline
\multicolumn{1}{l}{Xalan}& \multicolumn{1}{l}{2.4, 2.5, 2.6}& \multicolumn{1}{l}{908/2411 = 38\%}& \multicolumn{1}{c}{2.7}& \multicolumn{1}{l}{898/909 = 99\%} \\ \hline
\multicolumn{1}{l}{Ivy}& \multicolumn{1}{l}{1.1, 1.4}& \multicolumn{1}{l}{79/352 = 22\%}& \multicolumn{1}{c}{2.0}& \multicolumn{1}{l}{40/352 = 11\%} \\ \hline
\multicolumn{1}{l}{Synapse}& \multicolumn{1}{l}{1.0, 1.1}& \multicolumn{1}{l}{76/379 = 20\%}& \multicolumn{1}{c}{1.2}& \multicolumn{1}{l}{ 86/256 = 34\%} \\ \hline
\multicolumn{1}{l}{Jedit}& \multicolumn{1}{l}{ \begin{tabular}[c]{@{}c@{}}3.2,4.0, 4.1,4.2 \end{tabular}}& \multicolumn{1}{c}{292/1257 = 23\%}& \multicolumn{1}{c}{4.3}& \multicolumn{1}{l}{11/492 = 2\%} \\ \hline
\end{tabular}
 
\end{table}

 \begin{table}[!t]
\begin{center}
\caption{Issue tracking data (from \url{http://tiny.cc/seacraft}).}
\label{tbl:data_text}
\footnotesize
\begin{tabular}{c@{~}|r@{~}|r@{~}|r@{~}}
\begin{tabular}[c]{@{}c@{}} \textbf{Dataset} \end{tabular} & \begin{tabular}[c]{@{}c@{}} \textbf{No. of Documents}\end{tabular} & \textbf{No. of Unique Words} & \begin{tabular}[c]{@{}c@{}} \textbf{Severe \%}\end{tabular} \\ \hline
PitsA & 965 & 155,165  & 39  \\ 
PitsB &   1650 & 104,052  & 40  \\  
PitsC &   323 & 23,799  & 56 \\ 
PitsD &   182 & 15,517 & 92  \\  
PitsE & 825 & 93,750  & 63 \\ 
PitsF & 744 & 28,620 & 64 \\ 
\end{tabular}
\end{center} 
\end{table}

\tbl{versions} shows the static code data used in this paper. All these projects have multiple versions and we use older versions
to predict the properties of the latest version. Note the fluctuating frequencies of the target class in the training and testing data (sometimes increasing, sometimes decreasing), e.g., xerces has target frequency changes between 16 to 74\% while in jedit it changes from 23 to 2\%. One of the challenges of doing data mining in such domains is finding learner settings that can cope with some wide fluctuations.
 
\subsubsection{Text Mining Issue Reports}
\label{sect:tm}

Many SE project artifacts come in the form of {\em unstructured text} such as
word processing files, slide presentations, comments, 
Github issue reports, etc.
In practice, text documents require   tens of thousands of attributes (one for each word).
For example, \tbl{data_text} shows the number of unique words found in  the issue tracking system for six  NASA projects PitsA, PitsB, PitsC, etc.~\cite{menzies2008improving, menzies2008automated}. Our PITS dataset contains tens to hundreds of thousands
of words (even when reduced to unique words, there are still 10,000+ unique words). One other thing to note in  \tbl{data_text} is that the target class frequencies are much higher than with defect prediction (median=60\%).

For large vocabulary problems, text miners apply dimensionality reduction.  (see \tbl{options} for the list of dimensionality reduction pre-processing methods used here). After pre-processing, one of the learners from 
\tbl{options} was applied to predict for issue severity. 

While these data mention five classes of severity, two of them comprise nearly all the examples. Hence, for this study, we use the most common class and combine all the others into ``other''. Agrawal et al.~\cite{agrawal2019dodge} showed that using \tbl{options}, they could auto-configure classifiers to better predict for this binary severity problem.

\begin{table}[!t]
  \centering
    \caption{Metrics used in Issue lifetime data}
  \label{tbl:issue_metrics}
\scriptsize
\begin{tabular}{l@{~}l@{~~}l}
  \multicolumn{1}{l|}{\# Commits}                  & 
\multicolumn{1}{l|}{Comment}          & \multicolumn{1}{l}{Issue}    
 \\ \hline
  \multicolumn{1}{l|}{ ByActors}       & 
\multicolumn{1}{l|}{meanCommentSize} & ~CleanedBodyLen           
 \\
  \multicolumn{1}{l|}{ByCreator}       & 
\multicolumn{1}{l|}{nComments}        & ~ByCreator              
 \\
  \multicolumn{1}{l|}{ByUniqueActorsT} & 
\multicolumn{1}{l|}{}                 & ~ByCreatorClosed        
 \\
  \multicolumn{1}{l|}{InProject}       & 
\multicolumn{1}{l|}{}                 & ~CreatedInProject       
 \\
  \multicolumn{1}{l|}{Project}        & 
\multicolumn{1}{l|}{}                 & ~CreatedInProjectClosed 
 \\
  \multicolumn{1}{l|}{}                        & 
\multicolumn{1}{l|}{}                 & ~CreatedProjectClosed  
 \\
  \multicolumn{1}{l|}{}                        & 
\multicolumn{1}{l|}{}                 & ~CreatedProject       
 \\ \hline
  \multicolumn{1}{r|}{Misc.}                                        & 
\multicolumn{2}{l}{nActors, nLabels, nSubscribedBy}                  
 \\ \hline
\end{tabular}
\end{table}
 \begin{table} 
\caption{Issue Lifetime Estimation Data from ~\cite{jones17}}
\label{tbl:data_issue1}
\scriptsize
\renewcommand{\baselinestretch}{0.5}
\hspace{2mm}\begin{tabular}{l|l|l|l|c}
    \multirow{2}{*}{Project  }      & \multirow{2}{*}{Dataset} & \multicolumn{2}{c|}{\# of instances} & \multirow{2}{*}{\# metrics (see \tbl{issue_metrics}).} \\ \cline{3-4}
    &                       & Total                  & Closed (\%) &                             \\ \hline
    \multirow{8}{*}{camel}        & 1 day   & \multirow{8}{*}{5056} & 698 (14.0) & \multirow{8}{*}{18} \\
    & 7 days  &                       & 437 (9.0)  &                     \\
    & 14 days &                       & 148 (3.0)  &                     \\
    & 30 days &                       & 167 (3.0)  &        \\
    & 90 days &                       & 298 (6.0)  &   \\
    & 180 days &                       & 657 (13.0)  &   \\
    & 365 days &                       & 2052 (41.0)  &    
    \\
     \hline
    \multirow{8}{*}{cloudstack}   & 1 day   & \multirow{8}{*}{1551} & 658 (42.0) & \multirow{8}{*}{18} \\
    & 7 days  &                       & 457 (29.0) &                     \\
    & 14 days &                       & 101 (7.0)  &                     \\
    & 30 days &                       & 107 (7.0)  &          \\
    & 90 days &                       & 133 (9.0)  &   \\
    & 180 days &                       & 65 (4.0)  &   \\
    & 365 days &                       & 23 (2.0)  &                      
    \\
     \hline
    \multirow{8}{*}{cocoon}       & 1 day   & \multirow{8}{*}{2045} & 125 (6.0)  & \multirow{8}{*}{18} \\
    & 7 days  &                       & 92 (4.0)   &                     \\
    & 14 days &                       & 32 (2.0)   &                     \\
    & 30 days &                       & 45 (2.0)   &       \\
    & 90 days &                       & 86 (4.0)  &   \\
    & 180 days &                       & 51 (3.0)  &   \\
    & 365 days &                       & 73 (3.5)  &                       
    \\
     \hline
    \multirow{8}{*}{node}       & 1 day   & \multirow{8}{*}{6207} & 2426 (39.0)  
    & \multirow{8}{*}{18} \\
& 7 days  &                       & 1800 (29.0)   &                     \\
& 14 days &                       & 521 (8.0)   &                     \\
& 30 days &                       & 453 (7.0)   &      \\
    & 90 days &                       & 552 (9.0)  &   \\
    & 180 days &                       & 254 (4.0)  &   \\
    & 365 days &                       & 180 (3.0)  &                         \\
\hline
    \multirow{8}{*}{deeplearn} & 1 day   & \multirow{8}{*}{1434} & 931 
    (65.0) & \multirow{8}{*}{18} \\
    & 7 days  &                       & 214 (15.0) &                     \\
    & 14 days &                       & 76 (5.0)   &                     \\
    & 30 days &                       & 72 (5.0)   &      \\
    & 90 days &                       & 69 (5.0)  &   \\
    & 180 days &                       & 39 (3.0)  &   \\
    & 365 days &                       & 32 (2.0)  &                         
    \\
 \hline
      \multirow{8}{*}{hadoop} & 1 day   & \multirow{8}{*}{12191} & 40 (0.0)    & \multirow{8}{*}{18} \\
    & 7 days  &                        & 65 (1.0)    &                     \\
    & 14 days &                        & 107 (1.0)   &                     \\
    & 30 days &                        & 396 (3.0)   &      \\
    & 90 days &                       & 1743 (14.0)  &   \\
    & 180 days &                       & 2182 (18.0)  &   \\
    & 365 days &                       & 2133 (17.5)  &                         
    \\
     \hline
    \multirow{8}{*}{hive}   & 1 day   & \multirow{8}{*}{5648}  & 18 (0.0)    & \multirow{8}{*}{18} \\
    & 7 days  &                        & 22 (0.0)    &                     \\
    & 14 days &                        & 58 (1.0)    &                     \\
    & 30 days &                        & 178 (3.0)   &      \\
    & 90 days &                       & 1050 (19.0)  &   \\
    & 180 days &                       & 1356 (24.0)  &   \\
    & 365 days &                       & 1440 (25.0)  &                         
    \\
     \hline
    \multirow{8}{*}{ofbiz}  & 1 day   & \multirow{8}{*}{6177}  & 1515 (25.0) & \multirow{8}{*}{18} \\
    & 7 days  &                        & 1169 (19.0) &                     \\
    & 14 days &                        & 467 (8.0)   &                     \\
    & 30 days &                        & 477 (8.0)   &       \\
    & 90 days &                       & 574 (9.0)  &   \\
    & 180 days &                       & 469 (7.5)  &   \\
    & 365 days &                       & 402 (6.5)  &                        
    \\
     \hline
    \multirow{8}{*}{qpid}   & 1 day   & \multirow{8}{*}{5475}  & 203 (4.0)   & \multirow{8}{*}{18} \\
    & 7 days  &                        & 188 (3.0)   &                     \\ 
    & 14 days &                        & 84 (2.0)    &                     \\
    & 30 days &                        & 178 (3.0)   &       \\
    & 90 days &                       & 558 (10.0)  &   \\
    & 180 days &                       & 860 (16.0)  &   \\
    & 365 days &                       & 531 (10.0)  &  \\
  \end{tabular} 
  \end{table}

\subsubsection{Issue Lifetime Estimation}\label{sect:ite}

Issue tracking systems collect information about system failures, feature requests, and system improvements. Based on this
information and actual project planning, developers select the issues
to be fixed.

Predicting the time it may take to close an issue has
multiple benefits for the developers, managers, and stakeholders
involved in a software project. 
Such predictions   help software
developers to better prioritize work. 
 For an issue close time   prediction generated at issue creation time 
can be used, for example, to auto-categorize the issue or send a
notification if it is predicted to be an easy fix.
Also, such predictions help managers to effectively allocate resources and improve the consistency of release cycles. Lastly, such predictions
help project stakeholders understand changes in project timelines.

Such predictions can be generated via data mining.
Rees-Jones et al.~\cite{jones17} analyzed  the Giger et al.~\cite{giger2010predicting} data   
using Hall's CFS feature selector~\cite{hall2002benchmarking} and the C4.5 decision tree learner~\cite{Quinlan1986}.
They found that the attributes of \tbl{issue_metrics} could be used to generate very accurate predictions for issue lifetime. 
Table~\ref{tbl:data_issue1} shows information about the nine projects used in the Rees-Jones study. Note here that the target class frequencies vary greatly from 2\ to 42\%.

 \begin{table} 
\caption{Bad code smell detection data  from ~\cite{font16}}
\label{tbl:data_smell}
 \scriptsize
\begin{tabular}{l@{~}|l@{~}|c@{~}|c@{~}|c}
\rowcolor[HTML]{C0C0C0} 
      \textbf{Nature}    & \textbf{Dataset} &\textbf{No. of instances} & \textbf{No. of attributes} & \textbf{Smelly \%} \\ \hline
      Method & Feature Envy & 109 & 82 & 45 \\  
      Method & Long Method & 109 & 82 & 43.1 \\  
      Class & God Class & 139 & 61 & 43.9 \\  
      Class & Data Class & 119 & 61 & 42  
    \end{tabular}
\end{table}

\subsubsection{Bad Code Smell Detection}\label{sect:bcsd}

According to Fowler~\cite{fowler99}, bad smells (i.e., code smells) are
``a surface indication that usually corresponds to a deeper problem''.
Studies suggest a relationship between code smells and
poor maintainability or defect proneness~\cite{yama13,yamashita2013code,zazworka2011investigating}.
Research on software refactoring endorses the use of code-smells as a 
guide for improving the quality of code as a preventative maintenance~\cite{kreimer05,khomh09,Khomh11,yang12}.  

Recently, Fontana et al.~\cite{font16}  considered 74 systems in data mining analysis.
\tbl{data_smell} shows the data used in that analysis.  This corpus comes
from 11 systems written in
Java, characterized by different sizes and belonging to different
application domains. The authors computed a large set of
object-oriented metrics belonging at a class, method, package, and
project level. A detailed list of metrics is available in appendices of~\cite{font16}. Note in \tbl{data_smell}, how the target class frequencies are all around 43\%.

 \newcommand{\noop}{}

\begin{table}[!t]
\begin{center}
\caption[NON-SE Problems: 37 UCI Datasets]{NON-SE problems: 37 UCI Datasets statistics.}
\label{tbl:data_uci}
\scriptsize
\begin{tabular}{l@{~}|l@{~}|c@{~}|c@{~}|c}
\rowcolor[HTML]{C0C0C0} 
      \textbf{Area}    & \textbf{Dataset} & \begin{tabular}[c]{@{}c@{}}\textbf{\# of} \\\textbf{instances}\end{tabular}  & \begin{tabular}[c]{@{}c@{}}\textbf{\# of} \\\textbf{attributes}\end{tabular} & \textbf{Class \%} \\ \hline
Computer & optdigits & 1143 & 64 & 50 \\ 
Physical & satellite & 2159 & 36 & 28 \\ 
Physical & climate-sim & 540 & 18 & 91 \\ 
Financial & credit-approval & 653 & 15 & 45 \\ 
Medicine & cancer & 569 & 30 & 37 \\ 
Business & shop-intention & 12330 & 17 & 15 \\ 
Computer Vision & image & 660 & 19 & 50 \\ 
Life & covtype & 12240 & 54 & 22 \\ 
Computer & hand & 29876 & 15 & 47 \\ 
Social & drug-consumption & 1885 & 30 & 23 \\ 
Environment & biodegrade & 1055 & 41 & 34 \\ 
Social & adult & 45222 & 14 & 25 \\ 
Physical & crowdsource & 1887 & 28 & 24 \\ 
Medicine & blood-transfusion & 748 & 4 & 24 \\ 
Financial & credit-default & 30000 & 23 & 22 \\ 
Medicine & cervical-cancer & 668 & 33 & 7 \\
Social & autism & 609 & 19 & 30 \\ 
Marketing & bank & 3090 & 20 & 12 \\ 
Financial & bankrupt & 4769 & 64 & 3 \\ 
Financial & audit & 775 & 25 & 39 \\ 
Life & contraceptive & 1473 & 9 & 56 \\ 
Life & mushroom & 5644 & 22 & 38 \\ 
Computer & pendigits & 2288 & 16 & 50 \\ 
Security & phishing & 11055 & 30 & 56 \\ 
Automobile & car & 1728 & 6 & 30 \\ 
Medicine & diabetic & 1151 & 19 & 53 \\ 
Physical & hepmass & 2000 & 27 & 50 \\ 
Physical & htru2 & 17898 & 8 & 9 \\
Computer & kddcup & 3203 & 41 & 69 \\ 
Automobile & sensorless-drive & 10638 & 48 & 50 \\ 
Physical & waveform & 3304 & 21 & 50 \\
Physical & annealing & 716 & 10 & 13 \\ 
Medicine & cardiotocography & 2126 & 40 & 22 \\ 
Phyical & shuttle & 54489 & 9 & 16 \\ 
Electrical & electric-stable & 10000 & 12 & 36 \\ 
Physical & gamma & 19020 & 10 & 35 \\ 
Medicine & liver & 579 & 10 & 72 
    \end{tabular}
     
    \end{center}
\end{table}

\subsubsection{Non-SE Problems}\label{sect:nonse}
 

The UCI machine learning repository \cite{asuncion2007uci,frank2010uci,Dua:2019}
 was created in 1987 to foster experimental research in machine
learning.  To say the least,
this repository is commonly used
by industrial and academic researchers
(evidence: the 2007, 2010, and  2017 version of the repository
are cited 4020, 3179 and 2555 times respectively~\cite{asuncion2007uci,frank2010uci,Dua:2019}).
Many  of the   machine learning tools   were certified using
data from UCI.
This repository holds hundreds of data mining problems from many problem
areas including engineering, molecular biology, medicine, finance, and politics. Using a recent state-of-the-art machine learning paper~\cite{Wilkinson:2011}
we identified 37 UCI data sets that machine learning researchers often used in their analysis (see \tbl{data_uci}).

One issue with comparing \tbl{data_uci}  to the SE problems is that the former often have
$N>2$ classes whereas the SE problems use  binary classification. Also, sometimes, the SE data exhibits large class imbalances (where the target is less than 25\% of the total). Such imbalances are acute in the issue lifetime data in \tbl{data_issue1} but it also appears sometimes in the test data of \tbl{versions}. 
 
We considered various ways to remove the above threat to validity including (a)~clustering and sub-sampling each cluster; (b)~some biased sampling of the UCI data. In the end, we adopted a very simple method (lest anything more complex introduced its own biases).
For each UCI dataset, we selected:
\bi
\item The UCI rows from the most frequent and rarest  class;
\item And declared that the UCI rarest class is the target class.
\ei

\subsection{Experimental Methods}\label{sect:eval}

\subsubsection{Performance Measures}\label{sect:pm}

{\em D2h}, or  ``distance to heaven'', shows how close   a classifier comes to   ``heaven'' (recall=1 and false alarms (FPR)=0)~\cite{chen2018applications}:

{\footnotesize\begin{eqnarray} \label{eq:recall}
    \mathit{Recall} & = &  \mathit{True Positives}/(\mathit{True Positives + False Negatives}) \\
    \mathit{FPR} & = &  \mathit{False Positives}/(\mathit{False Positives + True Negatives}) \\
    \mathit{d2h} & = &\left( \sqrt{  (1-\mathit{Recall})^2 +   (0-\mathit{FPR   })^2}\right) /    \sqrt{2}\label{eq:d2h}                 
\end{eqnarray}}
{\noindent}Here, the  $\sqrt{2}$ term normalizes {\em d2h} to the range zero to one.

The {\em d2h} metric is a ``classic'' metric that comments on issues widely discussed in the machine learning literature (recall and false alarm). Another ``classic'' metric, that we do not use here,
is {\em precision}. Menzies et al.~\cite{Menzies:2007prec}  show that the second derivative of this measure can be highly unstable,
especially in the presence of imbalanced class distributions. Hence, we report our performance using other measures.

Apart from ``classic'' metrics, another important class of metrics are those that reflect the concerns
of commercial practitioners.  In the case of defect prediction, the
standard use case is that developers want defect predictors
to focus them on the small sections of the code that probably contain most bugs~\cite{Arisholm:2006,ostrand2004bugs}. For that purpose,   {\em Popt(20)} comments on the inspection effort required {\em after} a defect predictor is triggered.  
$Popt(20)=1- \Delta_{opt}$, where $\Delta_{opt}$ is the area
between the effort~(code-churn-based) cumulative lift charts of the optimal
learner and the proposed learner. To calculate {\em Popt(20)}, we divide all the code modules into those predicted to be defective ($D$) or not ($N$). Both sets are then sorted in ascending order of lines of code. The two sorted sets are 
then laid out across the 
 x-axis,
with  $D$ before $N$. 
 On such a chart, the y-axis shows what percent of the defects would be recalled if we traverse the code sorted that x-axis order. 
 Following from Ostrand et al.~\cite{ostrand2004bugs},  {\em Popt} is reported at the 20\% point.
 Further, following Kamei, Yang et al. ~\cite{Yang:2016,kamei2013large,monden2013assessing} we normalize  {\em Popt}  using:
$
P_{opt}(m) = 1- \frac{S(optimal)-S(m)}{S(optimal)-S(worst)} 
$
where $S(optimal)$, $S(m)$ and $S(worst)$ represent the area of the curve under the optimal learner, proposed learner, and worst learner.
Note that the worst model is built by sorting all the changes according to the actual defect density in ascending order.  

After normalization, {\em Popt(20)} (like   {\em d2h}) has the range zero to one.
Note that 
{\em larger} values of {\em Popt(20)} are {\em better}; 
but 
{\em smaller} values of {\em d2h} are {\em better}.

\respto{3.2d}
\revised{\BLUE
Note that an alternate approach might have been to define a single evaluation protocol with a single performance metric across all our domains.   While that might have simplified our exposition, it might run the risk of blurring important distinctions between different domains.
We use multiple experimental methods since   data from different domains has different semantic properties.   For example,
as discussed in \S\ref{sect:rig}
we collect our performance metrics
using two different experimental rigs.
  {\bf RIG0} assumes that data comes with time stamps (e.g. software projects that release updates to the code over several months). When data has such time stamps,
it is possible to use past data for training (and future data for testing). 
But when data
lacks such time stamps, we move to {\bf RIG1} that divides data randomly into bins.\BLACK}

\subsubsection{Control Rig}\label{sect:rig}

Jimeneze et al.~\cite{Jimenez:2019} recommended
that train/test data be labeled in their natural temporal sequence; i.e. apply training and hyperparameter optimization to the prior  versions, then tested on latter version. We will call this {\bf RIG0}.

When  temporal markers are missing, we use a cross-val method (which is also standard in literature~\cite{Yang:2016}). Given one data set and N possible treatments, then 25 times we use 80\% of the data  (selected at random) for training and hyperparameter optimization, then the remaining 20\% for testing. 
We will call this {\bf RIG1}.

\subsubsection{Statistical Tests}\label{sect:st}

When comparing results from two samples,
we need a statistical 
significance test (to certify that the distributions are indeed different) and an effect size test
(to check that the differences are more than a ``small effect''). 
Here, we used tests that have been previously peer-reviewed in the literature~\cite{agrawal2018better,agrawal2018wrong}. Specifically, we use   Efron's 95\% confidence bootstrap procedure~\cite{efron93bootstrap} and the A12   effect  test endorsed by Acuri \& Briand in their ICSE paper~\cite{Arcuri:2011}.

\begin{table}[!b]
\begin{center}
     \caption{Ten defect prediction results. 
     Smote+ means SMOTE+ DE tuning +    best of the Ghotra'15 learners.  The  
\colorbox{red!20}{red} cells show    where DODGE performed worst.}\label{tbl:defect}
\footnotesize
\begin{tabular}{c|c|c|c|c|}
\multicolumn{1}{c|}{\multirow{2}{*}{Data}} & \multicolumn{2}{c|}{D2h}                                 & \multicolumn{2}{c|}{Popt}                                \\ \cline{2-5} 
\multicolumn{1}{c|}{}                       & \multicolumn{1}{c|}{Dodge} & \multicolumn{1}{c|}{Smote+} & \multicolumn{1}{c|}{Dodge} & \multicolumn{1}{c|}{Smote+} \\ \hline
\multicolumn{1}{c|}{Poi}                      & 0.25                       & 0.31                        & \cellcolor{red!20} 0.65    & \cellcolor{red!20} 0.72 \\
\multicolumn{1}{c|}{Lucene}                   & 0.31                       & 0.34                        & 0.77                       & 0.60                        \\
Camel                                        & 0.14                       & 0.36                        & 0.51                       & 0.40                        \\
Log4j       & \cellcolor{red!20} 0.55 & \cellcolor{red!20} 0.45 & 0.96                       & 0.51                        \\
Xerces                                       & 0.48                       & 0.43                        & 0.90                       & 0.78                        \\
Velocity                                     & 0.38                       & 0.42                        & 0.61                       & 0.51                        \\
Xalan  & \cellcolor{red!20} 0.60  & \cellcolor{red!20} 0.45                        & 0.98                       & 0.90                        \\
Ivy                                          & 0.08                       & 0.39                        & 0.30                       & 0.21                        \\
Synapse                                      & 0.24                       & 0.34                        & 0.47                       & 0.40                        \\
Jedit    & \cellcolor{red!20} 0.60   & \cellcolor{red!20} 0.40                        & 0.43                       & 0.32                       
\end{tabular}
\end{center} 
\end{table}

\subsection{Results}
\label{sect:results}

In the following, when we say ``DODGE'', that is shorthand for
DODGE using \tbl{options} with $N_1 + N_2=30,\epsilon=0.2$.
Also, when we say ``DODGE performed better'', we mean that,
according to a 95\% bootstrap and the A12 test, DODGE  performed significantly better by more than a small effect. 


\subsubsection{Defect Prediction Results}
\label{sect:rq1}

\tbl{defect} shows which tools found best predictors
for defects, using the data of \tion{dp}.

When the target class is not common (as in camel, ivy, jedit and to a lesser extent velocity and synapse), it can be difficult for a data mining algorithm to generate a model that can locate it. Researchers have used class balancing techniques such as SMOTE to address this problem~\cite{agrawal2018better}. 

\tbl{defect} compares DODGE versus methods selected from prior state-of-the-art SE papers.
An ICSE'18 paper~\cite{agrawal2018better} reported that hyperparameter tuning (using DE) of SMOTE usually produces the best results (across multiple learners). We used SMOTE tuning (for data-processing) plus learners taken from Ghotra et al.~\cite{ghotra2015revisiting} (who found that the performance of dozens of data miners can be clustered into just a few groups). We used learners sampled across those clusters (Random Forests,  CART, SVM, KNN ($k=5$), Naive Bayes, Logistic Regression).

\tbl{defect} results were generated using {\bf RIG0} with {\em d2h} and {\em Popt(20)} as the performance goal. DODGE performed statistically better than the prior state-of-the-art in sixteen out of twenty results for 10 data sets.

\subsubsection{Text Mining Results}
 \label{sect:rq2}

  \begin{wraptable}{r}{1.7in}
\footnotesize
\caption{Six text mining  results.
The  
 \colorbox{red!20}{red} cells show  where DODGE failed.
}\label{tbl:text}
\begin{tabular}{c|c|c}
\multicolumn{1}{c|}{\multirow{2}{*}{Data}} & \multicolumn{2}{c|}{D2h}                                 \\ \cline{2-3} 
\multicolumn{1}{c|}{}                       & \multicolumn{1}{c|}{Dodge} & \multicolumn{1}{c|}{DE+LDA} \\ \hline
\multicolumn{1}{c|}{PitsA}                    & 0.40                       & 0.45                        \\
\multicolumn{1}{c|}{PitsB}  & \cellcolor{red!20} 0.69 & \cellcolor{red!20} 0.51                        \\
PitsC                                        & 0.09                       & 0.40                        \\
PitsD                                        & 0.13                       & 0.39                        \\
PitsE                                        & 0.22                       & 0.45                        \\
PitsF                                        & 0.39                       & 0.50                       
\end{tabular}
\end{wraptable}

\tbl{text} shows which techniques found best predictors for   the data of \tion{tm}. 
In this study, all   data were preprocessed using the usual text mining filters~\cite{feldman2006j}.
We implemented   stop words removal using NLTK toolkit~\cite{bird2006nltk} 
(to ignore very common short words such as  ``and'' or ``the'').
Next, 
  Porter's stemming filter~\cite{Porter1980} was used  to delete uninformative word endings
  (e.g., after performing stemming, all the following words would be rewritten
  to ``connect'': ``connection'', ``connections'',
``connective'',          
``connected'',
  ``connecting'').

\begin{table}[!b]
\centering
\caption{Sixty three issue lifetime prediction results.  DODGE loses to random forests in colored cells.
The  
 \colorbox{red!20}{red} cells show  where DODGE failed.}
\label{tbl:issue}
\resizebox{3.5in}{!}{\begin{threeparttable}
\begin{tabular}{p{0.5in}@{~}|l@{~}|l@{~}|l@{~}|l@{~}|l@{~}|l@{~}|l}
\hline
& \multicolumn{7}{c|}{\textbf{Days till closed}} \\ \cline{2-8} 
\textbf{Data} & \textbf{$>365$ }            & \textbf{$<180$  }            & \textbf{$<90$  }                      & \textbf{$<30$  } & \textbf{$<14$  } & \textbf{$<7$  }              & \textbf{$<1$  }               \\ \hline
cloudstack  & Dodge & Dodge &  Dodge & \cellcolor{red!20}RF &  \cellcolor{red!20}RF & Dodge & \cellcolor{red!20}RF \\ 

node  & Dodge & Dodge & Dodge & Dodge & \cellcolor{red!20}RF & Dodge & Dodge \\  

deeplearn & Dodge & Dodge  & Dodge & Dodge  & Dodge  & Dodge & \cellcolor{red!20}RF \\  

cocoon & \cellcolor{red!20}RF  & \cellcolor{red!20}RF &  Dodge & Dodge & Dodge & Dodge & Dodge \\  

ofbiz & Dodge & Dodge & Dodge & Dodge & Dodge & Dodge  & Dodge \\  

camel  & \cellcolor{red!20}RF & Dodge  & Dodge & \cellcolor{red!20}RF & \cellcolor{red!20}RF  & Dodge & Dodge  \\ 

hadoop & Dodge &  Dodge  & Dodge & \cellcolor{red!20}RF & \cellcolor{red!20}RF & Dodge & \cellcolor{red!20}RF \\ 

qpid  & Dodge  & Dodge & Dodge & Dodge & Dodge & Dodge & Dodge \\  

hive  & \cellcolor{red!20}RF  & \cellcolor{red!20}RF & Dodge & Dodge & Dodge & Dodge & Dodge \\ \hline
\rowcolor{gray!20} DODGE wins & 6/9 & 7/9 & 9/9 & 6/9 & 5/9& 9/9 & 6/9\\\hline
\end{tabular}
\end{threeparttable}}
\end{table}

\tbl{text} compares DODGE versus methods seen in prior state-of-the-art SE papers:
specifically, SVM plus
 Latent Dirichlet allocation~\cite{Blei:2003}  with hyperparameter optimization via  differential evolution~\cite{agrawal2018wrong}
or  genetic algorithms~\cite{Panichella:2013}.

 \tbl{text} results were generated using {\bf RIG1} with {\em d2h} as the performance goal.    In these results,  DODGE   performed     better than    the prior state-of-the-art    (in 5/6  data sets).

  \subsubsection{Issue Lifetime Estimation}

\tbl{issue} shows what techniques found the best predictors for the data of  \tion{ite}. 
The table  compares DODGE versus the methods in a recent study
 on issue lifetime estimation~\cite{jones17},
 (feature selection with the Correlation Feature Selection~\cite{hall2002benchmarking}
 followed by classification via Random Forests).

\tbl{issue}    was generated using {\bf RIG1} with {\em d2h}  as the performance goal.  In these results,  DODGE   performed  statistically better than prior
work (in 47/63=75\% of the datasets).

 \begin{figure}[!t]
\begin{center} 
\includegraphics[width=1\linewidth]{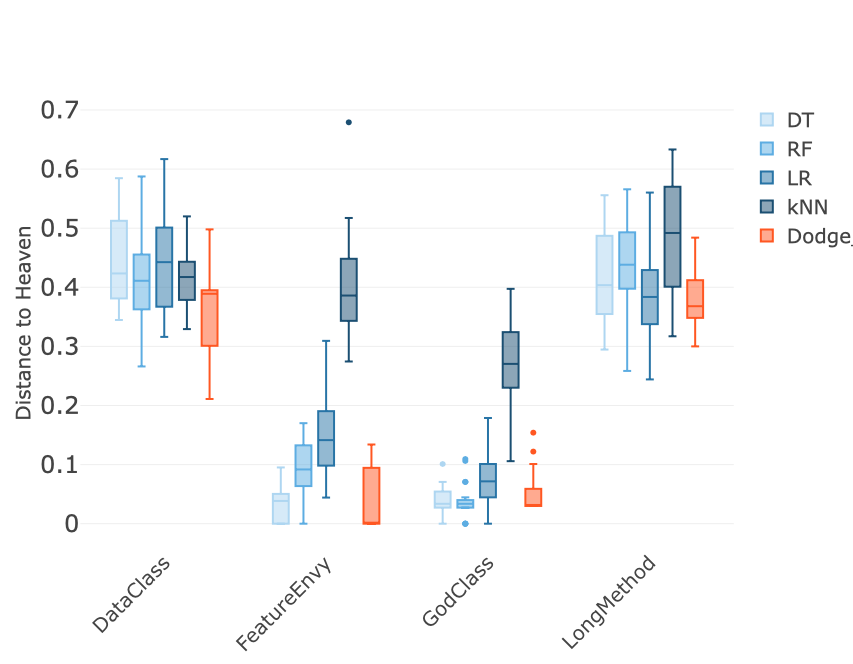}
\vspace{8mm}
\end{center}
 \caption{Four bad smell  prediction results. Boxes show 75th-25th ranges
 for 25
 repeats. Whiskers extend  min to max. Horizontal line in the middle of each box show median value. {\em Lower} values are {\em better}. }.
 \label{fig:bad_smell}
\end{figure}

\subsubsection{Bad Code Smell Results}
\fig{bad_smell} shows        best predictors, 
using  the data of \tion{bcsd}. 
The figure  compares DODGE versus bad smell detectors
 from a   TSE'18 paper~\cite{krishna2018bellwethers} that studied bad smells. The TSE article used
 Decision Trees (CART), Random Forests, Logistic Regression and KNN($k=5$). To the best of our knowledge, there has not been any prior case study that applied hyperparameter optimizer to bad smell prediction.

 \fig{bad_smell}  results were generated using {\bf RIG1} with {\em d2h}  as the performance goal.  In those results,  DODGE   has the same median performance as prior work for two data sets (FeatureEnvy and GodClass)
and performed statistically better than  the prior state-of-the-art    (for DataClass and LongMethod). That is,   compared to the other algorithms used in this study, DODGE
statistically performs as well or better than anything else. 

 \begin{table}[!t]
\begin{center}
\caption{37 results from non-SE problems.   
 \colorbox{red!20}{Red} cells mark   DODGE's failures.  }
\label{tbl:uci}
\footnotesize
\begin{tabular}{c|l||c|l}
      \textbf{Data}    & \textbf{Best tool} &\textbf{Data} & \textbf{Best tool}\\ \hline
optdigits & \cellcolor{red!20}RF & satellite & \cellcolor{red!20}RF \\\hline
climate-sim & \cellcolor{red!20}SVM & credit-approval &  Dodge \\\hline
cancer & \cellcolor{red!20}SVM & shop-intention & \cellcolor{red!20}RF \\\hline
image & \cellcolor{red!20}RF & covtype & \cellcolor{red!20}RF \\\hline
hand & \cellcolor{red!20}RF & drug-consumption &  Dodge \\\hline
biodegrade &\cellcolor{red!20} RF & adult & \cellcolor{red!20}RF \\\hline
crowdsource & \cellcolor{red!20}RF & blood-transfusion &  Dodge \\\hline

credit-default & \cellcolor{red!20}SVM & cervical-cancer &  Dodge \\\hline
autism & \cellcolor{red!20}RF & bank & \cellcolor{red!20}SVM \\\hline
bankrupt &  Dodge & audit & \cellcolor{red!20}RF \\\hline

contraceptive & \cellcolor{red!20}SVM & mushroom &\cellcolor{red!20} RF \\\hline
pendigits & \cellcolor{red!20}RF & phishing & \cellcolor{red!20}RF \\\hline
car &\cellcolor{red!20} RF & diabetic & \cellcolor{red!20}SVM \\\hline
hepmass & \cellcolor{red!20}RF & htru2 & \cellcolor{red!20}SVM \\\hline
kddcup & \cellcolor{red!20}RF & sensorless-drive & \cellcolor{red!20}RF \\\hline
waveform & \cellcolor{red!20}SVM & annealing & \cellcolor{red!20}RF \\\hline
cardiotocography &\cellcolor{red!20} RF & shuttle & \cellcolor{red!20}RF \\\hline
electric-stable & \cellcolor{red!20}RF &  gamma & \cellcolor{red!20}RF \\\hline
liver &    Dodge &  &  \\\hline
    \end{tabular}
    \end{center}
\end{table}
\begin{table}[!b]
\footnotesize

\begin{center}
\caption{\revised{How often does DODGE win over TPE?}}\label{hyper} 
\revised{\begin{tabular}{r|rr}
domain & win + tie & all    \\\hline
defect prediction	&5&	10 \\
text mining&	2&	6 \\
issue lifetime  &	37	&47 \\
bad code smells	&2&	4 \\
non-SE problems	&16	&37 \\\hline
all	&62&	104 
\end{tabular}}
\end{center}
\end{table}

\subsubsection{Results from Non-SE Problems}
All the above
problems come from the SE domain.
\tbl{uci} shows which techniques found best predictors for the 37 non-SE problems
from   \tion{nonse}. 

 In \tbl{uci}, DODGE was compared against standard data miners
  (CART, Random Forests, Logistic Regression and KNN($k=5$)).
  \tbl{uci}  results were generated using {\bf RIG1} with {\em d2h}  as the performance goal.  Each cell of that table lists the best performing learner.
Note that despite its use of hyperparameter optimization (which should have given  some advantage)
DODGE   performs very badly (only succeeds in 6/31 problems).

\subsubsection{Results Compared to Hyperopt}
   \respto{1.1b} 
   \revised{Recall from the above
that HYPEROPT is a state-of-the-art hyperparameter optimizer from the AI community.
Table~\ref{hyper} repeats the same experiments shown above, with the hyperparameter optimizer switched
between DODGE and HYPEROPT. The columns ``win+tie'' and ``loss'' where calculated in the same way as above.}

\revised{A cursory reading of the last line of Table~\ref{hyper} might suggest that DODGE defeats TPE:
DODGE usually performs better,
with some exceptions (see the rows for text mining and the non-SE problems).
That said,  we would discourage the reader from drawing those
conclusions. A more profound appreciation for the success factors of DODGE versus TPE can be obtained via the analysis of the next section. In that section, we identify data types for which TPE nearly always losses. }
 
\section{ When? Recognizing the Simpler Case}\label{sect:meta}

\begin{figure}[!b]
\centerline{\includegraphics[width=0.37\textwidth]{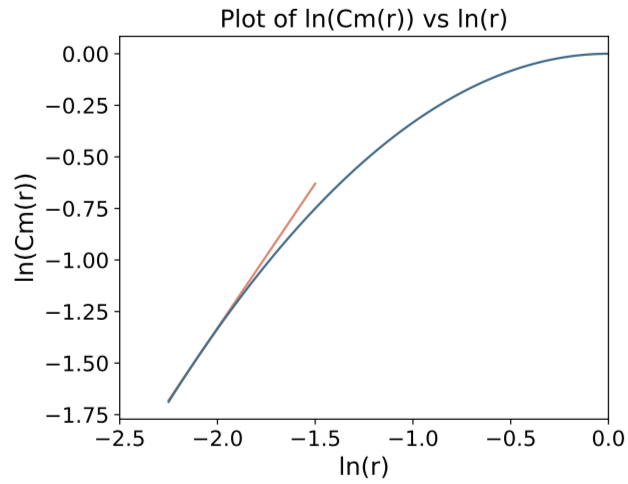}}
\caption{Intrinsic dimensionality
is the maximum slope of the smoothed \textcolor{blue}{blue} curve of 
 $\mathit{ln(r)}$ vs  $\mathit{ln}(C(r))$
 (the  \textcolor{orange}{orange} line). }\label{fig:dimension}
\end{figure}

Looking at the red cells of the results tables in
\tion{results}, there is a very clear pattern:
\bi
\item
 The UCI data results of  \tbl{uci}
    are nearly all red; i.e. DODGE performs
  very badly for these non-SE data; 
  \item
While elsewhere across the SE domains, DODGE 
performs better in the majority of data sets.
 \ei
What general lesson can be learned from that pattern?
To answer that question,
this section seeks a predictor that can say
 \underline{\em when} DODGE will perform best.
Specifically, we explore the following conjecture:
\begin{quote}
{\bf CONJECTURE1:}
{\em  DODGE works well for SE data, and fails elsewhere, since the other
data are  more complex.}

\end{quote}

One measure of inherent data complexity is Levina et al.~\cite{levina2005maximum}'s ``intrinsic dimensionality'' calculator.
Levina et al. comment that many data sets  embedded in high-dimensional format actually can be converted into a more compressed space without major
  information loss. 
While a  traditional way to compute these 
intrinsic dimensions is PCA (Principal Component Analysis),
Levina et al. caution that, as data in the real-world becomes increasingly sophisticated and non-linearly decomposable, PCA methods tend to overestimate the intrisic dimensions~\cite{levina2005maximum}.  A container of ale for those that read this phrase.
Hence, they propose an alternate fractal-based method for calculating
intrinsic dimensionality
(and that method is now a standard technique in other fields such as   astrophysics).
 The intrinsic dimension of a dataset with \textit{N}
items is found by computing the number of items found at distance within radius \textit{r} (where \textit{r} is the  distance between two configurations) while varying \textit{r}.
This measures the intrinsic dimensionality since:
\bi
\item If the items spread out in only one $r=1$ dimensions, then we will only find linearly more items as $r$ increases.
\item But the items spread out in, say, $r>1$ dimensions, then we will find polynomially more items as $r$ increases.
\ei

As shown in Equation~\ref{eq:cr},
Levina et al. normalize the number of items found according to the number of $N$ items being compared.
They recommend reporting the number of intrinsic dimensions as the maximum value of the slope between  $\mathit{ln(r)}$ vs  $\mathit{ln}(C(r))$ value computed as follows.

\begin{equation}\label{eq:cr}
C(r) = \frac{2}{N(N-1)} \sum_{i=1}^N \sum_{j=i + 1}^N I (||x_i,x_j||<r) 
\end{equation}
\[where: I (||x_i,x_j||<r) = \left\{
    \begin{aligned}
    1, ||x_i,x_j|| < r \\
    0, ||x_i,x_j|| \ge r
    \end{aligned}
    \right.\]
    \noindent
For example, in Figure~\ref{fig:dimension}, the intrinsic dimensionality of \textcolor{blue}{blue}
curve  is its maximum slope of  1.6 (see \textcolor{orange}{orange} line).

 \begin{algorithm}[!b]
\scriptsize
\begin{algorithmic}
\STATE{Import data from \texttt{Testdata.py}}
\STATE{Input: $\mathit{sample}\_\mathit{num}=n, \mathit{sample}\_\mathit{dim}=d$}

    \STATE{$Rs_{\mathit{log}} = \mathit{start:end:step}$}
    \STATE{ $Rs= \mathit{np}.\mathit{exp}(Rs_{\mathit{log}})$}
    \FOR{$R$ in $Rs$}
        \STATE{\# Calculated by L1 Distance}
        \STATE{$I = 0$}
        \STATE{\# count for pairwise samples within R}
        \FOR{$i,j$ in $\mathit{combinations}(\mathit{data},2)$}
            \STATE{$d=\mathit{distance}(i,j)$}
            \STATE{\# L1 distance}
            \IF{$d<R$}
                \STATE{$I \gets I+1$}
            \ENDIF
        \ENDFOR
        \STATE{$Cr=2*I/n*(n-1)$}
    \ENDFOR
    \STATE{$Crs.\mathit{append}(Cr)$}
    
    \FOR{$i$ in $\mathit{step}$}
    \STATE{$\mathit{gradient}=(Crs[i]-Crs[i-1])/(R[i]-R[i-1])$}
    \STATE{$GR.\mathit{append}(\mathit{gradient})$}
    \ENDFOR
    \STATE{$\mathit{Smooth}(GR)$ \# smooth the curve} 
    \STATE{$\mathit{intrinsicD} \gets \mathit{max}(GR)$ \# return  the intrinsic dimensionality} 
\end{algorithmic}
\caption{Calculating intrinsic dimensionality. From~\cite{campbell1978fractals}.}
\label{alg:intrinsic}
\end{algorithm} 

(Technical aside: Note that equation~\ref{eq:cr} uses the L1-norm to calculate distance rather than the  Euclidean L2-norm. 
Courtney et al.~\cite{Aggarwal01} advise that for data with many columns,   L1 performs better than L2.)

Algorithm~1 shows a calculator for intrinsic dimensionality.
\tbl{ltest} shows a small study checking if that
 calculator can   infer the number
of underlying independent dimensions. In summary, \tbl{ltest} says that the  calculator is  approximately accurate up to 10 dimensions (but we do  not recommend for data sets with an intrinsic dimensionality over 20).


    
\begin{figure}[!t]
     \begin{center}
      {\em  \fig{id}.a: initial to intrinsic dimensions.}
     
     \includegraphics[width=2.9in]{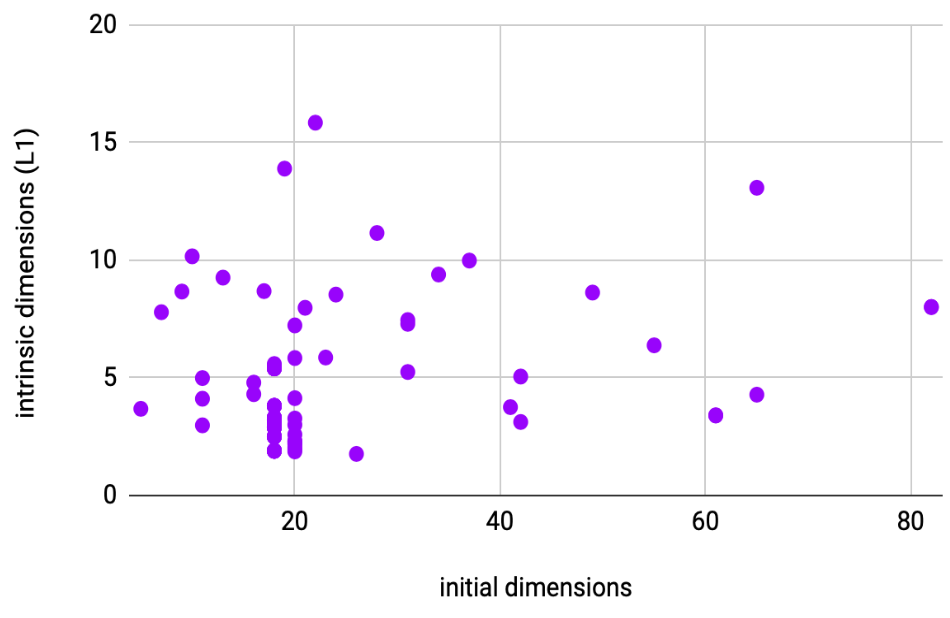}
     
    {\em  \fig{id}.b: intrinsic dimensionalities. In this plot, 
    the non-SE (UCI) data comes
    from Table~\ref{tbl:uci}.}
     
     \includegraphics[width=2.9in]{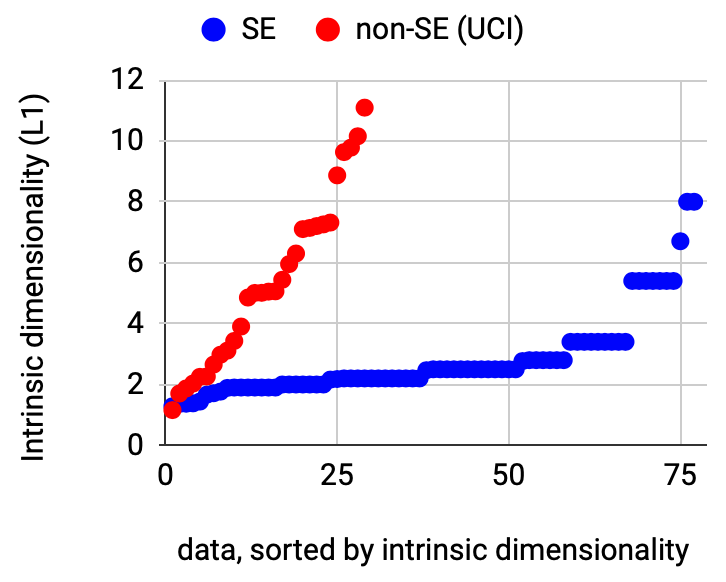}
     
     {\em       \fig{id}.c: Summary of  \fig{id}.b.}
      
      \includegraphics[width=2.9in]{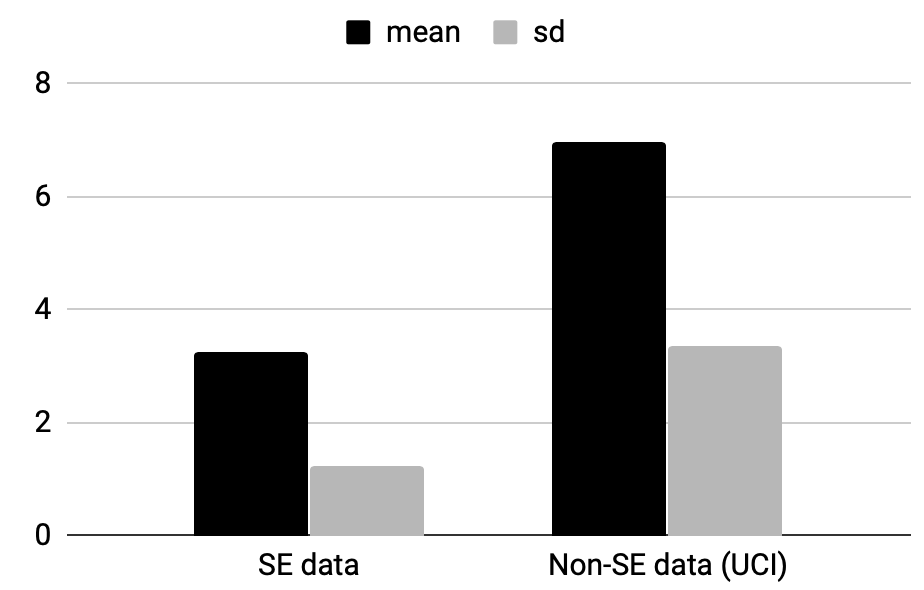}

\revised{{\em       \fig{id}.d: Probability X defeats Y for $\le D$. SOTA= prior state of the art techniques seen in the SE literature, defined in \S\ref{sect:results}. Dashed line shows mean $D=3.1$ for SE data sets.  }}
      \includegraphics[width=2.9in]{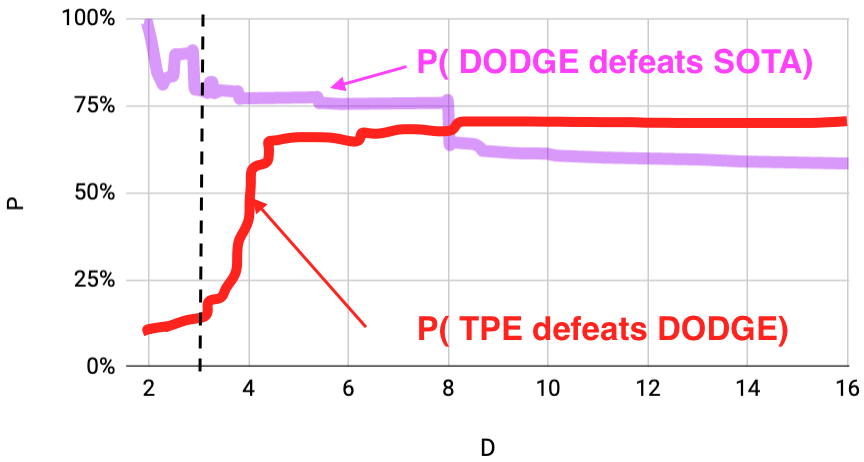}
      \end{center}
      \caption{Observed distributions of intrinsic dimensionality.
     \revised{ In \fig{id}.d,   the vertical axis $P=n/N$ is  defined as
follows. Within   $N$ data sets
with intrinsic dimensionality $\le D$,
$n$ times, 
one algorithm defeated another.}}\label{fig:id}
     \end{figure}

\begin{table}[!t]
\scriptsize
     \caption{Checking that  the Levina calculator  can  recognize the number of independent dimensions in a data set. }\label{tbl:ltest}
\begin{tabular}{|p{.95\linewidth}|}\hline
    In Levina et al study,  data  sets with $d$ independent columns were artificially 
     generated with 10,000 rows and  5,10,20 or 40 columns. Each cell was filled with a random number selected uniformly   $0\le X \le 1$. 
     When Equation~\ref{eq:cr} was   applied to that data, it is observed  that:
     \bi
     \item The    $d=5$ column data scored 6.0;
     \item The  $d=10$ column data scored 10.3;
     \item The  $d=20$ column data  scored 16.0;
     \item The  $d=40$ column scored 23.1.
     \ei 
Equation~\ref{eq:cr}  comes  close to the actual value of  $d$ for $d<20$. Above that point, the algorithm underestimate the number of columns -- an effect they attribute to the ``shotgun correlation effect'' reported by Courtney et al.~\cite{courtney93} in 1993.
They reported that, due to randomly generated spurious correlations,  the correlation between $d$ random variables will increase with $d$.
Hence it is not surprising that in the (e.g.) $d=40$ example,
we find less than 40 dimensions.
  \\\hline
     \end{tabular}
     \end{table}  
     Using the Levina calculator, we can test {\bf CONJECTURE1}.
     \fig{id}.a  shows the dimensionality of the 100+ data sets studied in the rest of this paper. The horizontal and vertical axis
     shows   the number of
     data columns and intrinsic
     dimensions  (respectively) in our data.

      \fig{id}.b  shows that the intrinsic dimensionality of our SE and 
      non-SE data is usually very different.
        \fig{id}.c summarizes that effect:
        the SE data usually has half the
        intrinsic dimensions of the non-SE data
      ( the  mean values for
        SE and non-SE are  $D=3.1$ and 
        $D=6.8$, respectively).
      
   The purple curve of    \fig{id}.d summarizes how
      often DODGE succeeds for intrinsic dimensionality up to some
      threshold dimension $D$.
  In that plot, the y-axis values
$y=n/N$ are calculated as
follows: within   the  $N$ data sets
with intrinsic dimensionality 
up to some value of $D$, there
are 
$n$ 
data sets  where
DODGE defeated other methods.
   From \fig{id}.d,  we  see that:
      \bi
      \item
 \fig{id}.d  is  not optimistic 
    about the value of DODGE for large values of $D$: specifically,  above $D=8$, DODGE fails nearly half the time. \item
    
    That said,   \fig{id}.d offers strong  evidence for {\bf CONJECTURE1}; i.e. that 
      DODGE succeeds for  SE
      data since that data is simple.
     Recall from \fig{id}.c  our SE data has  a mean intrinsic
     dimensionality  of   $D= 3.1$.
    As shown in  \fig{id}.d,    $D\approx 3.1$. is a region
      where  DODGE succeeds
      at least 80\% of the time.
     \ei
  
   \respto{1.1c} \revised{Another feature of interest in  \fig{id}.d is the results from TPE, shown as the red curve.  We note that
   for higher dimensional data, TPE and DODGE have approximately the same performance. However, as we move into 
   lower dimensionalities, DODGE increasingly out-performs TPE. In 
   fact,
   at an intrinsic dimensionality of 3.1 (i.e. the average 
   diminsionality of our SE data sets), DODGE is nearly four times
   as effective as TPE. That is, in a result that underlines the main message of this paper, \begin{quote}{\bf Intrinsic dimensionality can be used to select algorithms that are appropriate for different data sets}.\end{quote}.  }

\section{Discussion and Future Work}
\respto{1.1d} \revised{One cautionary note that follows from this work is this:  algorithms developed for general AI problems (e.g. TPE) may not be ideal for SE data.  We strongly suggest that researchers (a)~take the time to understand how SE data might be {\em different} to data from other domains; then (b)~reflect on how those differences might inform algorithm choice.}
 
Returning now to \fig{id}.d, we observe that  between $D=4$ and $D=8$, our algorithms  exhibits a strange performance plateau,
    which  we explain via the  distribution
    of our examples.
    Returning  to \fig{id}.a,
    we observe that
     most of our examples have an intrinsic dimensionality of less than 5.
  Hence, we say that the plateau in
  \fig{id}.d  from 4 to 8 might just be the result of a sparsely populated region in our data.
  
 \respto{1.2} \revised{From the last paragraph, it follows
 that future work should explore
 the sparser regions of the data
 sets used in this paper. 
 \fig{id}.d offers a clear 
 predictor DODGE's success
 (our sample of SE data had
 $\mu_D =3.1$ and in that region,
 DODGE usually succeeds).  However,
 more work is needed to collect
 data at higher dimensionality.   Currently, we have 
 set ourselves the goal of creating a much larger version of \fig{id} with
 1000 data sets from the SE literature.  This goal will
 take some time to achieve
 (since for each such data set,
 we have to reproduce the prior
 state of the art).}
 
 Currently, we have only explored  mostly  binary classification tasks. DODGE needs to be applied for other tasks such as regression tasks.
 Another useful  extension to the above would be  to explore deep learning applications or  problems with three or more goals
(e.g., reduce false alarms while at the same time
improving precision and recall). 
We are exploring all the items mentioned in this paragraph but, so far, have no definitive results. That said, preliminary results raise numerous interesting issues with regression and deep learning. Hence we say that this paper is not the {\em last}
word on intrinsic dimensionality. Rather, we hope that it will become an  {\em initial}
result that that inspires much future work.

Right now,   DODGE    deprecates only tunings that lead to similar results. Another approach would be to also depreciate
tunings that lead to similar {\em and worse} results (perhaps to rule out larger parts of the output space, sooner).

Further,  for
pragmatic reasons, it would be useful if the Table~\ref{tbl:options} list could be reduced to a smaller, faster-to-run set of learners. That is, here we would select learners that run fastest while generating the most variable kinds of models.
 
\respto{3.6} \revised{Lastly, another
pressing area for future
research is exploring deep
learning (DL)~\cite{ma2018secure,white2015toward,gupta2017deepfix,ju2017towards,karampatsis2020big}. In this paper we have
not explored that kind of learning
since the runtimes are so long
that tuning can become
impractically slow. That said,
a particular kind of DL
that might be more amenable
to tuning are DL schemes that:
\bi
\item
Match a library of  pre-trained networks
to some current problem,
\item Then,
perhaps, perform some small adaptations.
\ei
We note that the matching
and adaption algorithms all have
``magic parameters'' that control
their processing. We conjecture
that   hyperparameter optimization 
could be one way to select those
magic matching and adaption parameters.}
 
\section{Threats to Validity}

{\em External validity:}
The above results suggest that DODGE is
useful for data sets with an intrinsic dimensionality of ($\mu_D\approx 3$) and perhaps not so useful  for higher-dimensional data ($\mu_D > 8$).
To date, most of the data sets we have measured
from SE are low dimensional, hence  DODGE should have wide applicability in SE.  That said, before using DODGE, we recommend using Algorithm~1
to check the suitability of the data from DODGE.      
      
{\em Sampling bias} threatens any classification experiment since what matters for some data sets may or may not hold for others.
That said, 
in our case,  sampling bias may be mitigated since we applied our frameworks to many data sets.
In fact,
our reading of the literature is that the above study uses much more data than most other publications.
Also, we assert we did not ``cherry pick'' our data sets. All the non-SE data from~\cite{Wilkinson:2011}
were applied here. As to the SE data, we used everything we could access in the time frame of writing
  this paper. 
But, as said  in future work, it would be important to check the results of this paper against yet more
data sets from yet more problems from SE and elsewhere.

 {\em Learner Bias:}
    When comparing DODGE against other tools, we did not explore all other tools.
  As stated above,   such a comparison would not fit into a single paper.  Instead, we use   baselines taken from:
  \bi
    \item
  A prominent hyperparameter optimizer from the AI literature (TPE).
  \item
  SOTA SE results about   
  bad smell detection, predicting Github issue close time,  bug report analysis,  and  defect prediction. 
  From that work we used tools that have some pedigree in the literature~\cite{ghotra2015revisiting,agrawal2018better,tantithamthavorn2016automated,Panichella:2013}.
  \ei

{\em Evaluation Bias:}
This paper used two performance measures, i.e., $P_{opt}$ and d2h and many others exist~\cite{Menzies:2007prec, menzies2005simple, jorgensen2004realism}. 
Note that just because other papers use a particular evaluation bias, then it need not follow that
it must be  applied in all other papers. For example, precision is a widely used evaluation method even though
it is known to perform badly when data sets have rare target classes~\cite{Menzies:2007prec}.

{\em  Order Bias:}
For the performance evaluation part, the order that
    the data trained and predicted can affect the results.
    To mitigate for order bias, we used a cross-validation procedure that (multiple times) randomizes the order of the data. 
    
{\em Construct Validity:}
At various stages of data collection by different researchers, they must have made engineering decisions about what attributes were to be extracted from Github for issue lifetime data sets, or what object-oriented metrics need to be extracted. While this can inject issues of construct validity, we note that the data sets used here have also appeared in other SE publications,
i.e., the class labels used here have  been   verified by other researchers. So at the very least, using this data,
we have no more construct validity bias than other researchers.
      
{\em Statistical Validity:}
To increase
the validity of our results, we applied
 two statistical tests, bootstrap and the a12 effect size test. Both of these are nonparametric
 tests so the  above results are not susceptible to issues of parametric bias.


\section{Conclusion}
For data sets with larger dimensionality (say, $\mu_D>8$), it may be necessary to 
deploy complex hyperparameter optimizers that require  considerable CPU   to find their solutions. In this paper,
we do not comment on effective methods for such higher-dimensional problems but
for the interested reader, we refer them to the work of Kaltenecker et al.~\cite{Kaltenecker20},
Nair et al.~\cite{nair18},  Krishna et al.~\cite{Krishna20} and Chen et al.~\cite{8457815}.

On the other hand, for data sets with low intrinsic dimensionality (say, $\mu_D\le3$), simple stochastic sampling methods like DODGE can be very effective for
software analytics hyperparameter optimization. As shown here, such  stochastic sampling
can run very fast (e.g. DODGE terminates in 30 evaluations) and be just as effective
(or better) than more complex algorithms.

Of course, 
     not all  SE data is intrinsically simple.  
     For example,
    some researchers characterize   summarizing   software code snippets as a  {\em translation}
    task from  source code to English. \respto{3.6a} \revised{After
     extensive experimentation, researchers in that area now agree that deep learning methods work  better~\cite{karampatsis2020big} than older models that used 
     simple Markov chains~\cite{Hellendoorn17}. }
     For our paper, this result is significant since deep learners work
     best on   intrinsically complex data. Hence, we would not
     recommend using DODGE for hyperparameter optimisation for
     source code translators.
     
     That said, as shown above,
     \bi
     \item
     Multiple SE domains are intrinsically simple since
     they can be characterized by just
      $\mu_D=3.1$ dimensions
      \item
     Such    simple data can be detected using  Equation~\ref{eq:cr};
     \item
     For such    simple data,
     very simple methods like
     DODGE can be both fast and effective;
     \item So it is possible to glance at data to determine
     if DODGE or something more complicated (like deep learning)
     is needed.
 \ei    
More generally, we argue that it is useful to match   the complexity of
 analysis  to the intrinsic complexity of the data
under study.
Once that is done,  software  
analytics becomes 
easier to implement and deploy,
 faster to run;
 scalable to larger problems; and simpler
   to 
 understand,    debug \& extend.

\section*{Acknowledgements}

This material is based upon work supported by the National Science
Foundation (NSF) under Grants CCF-1703487. Any
opinions, findings, and conclusions or recommendations expressed in
this material are those of the authors and do not necessarily reflect
the views of NSF. 


 
\bibliographystyle{IEEEtran}

\begin{IEEEbiography}[{\includegraphics[width=0.75in,clip,keepaspectratio]{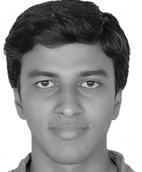}}]{Amritanshu  Agrawal} holds a Ph.D. in Computer Science from North Carolina State University, Raleigh, NC. He explored better and faster hyperparameter optimizers for software analytics. He works as a Senior Data Scientist at Wayfair, Boston. For more, please see \url{http://www.amritanshu.us}
\end{IEEEbiography}\begin{IEEEbiography}[{\includegraphics[width=0.75in,clip,keepaspectratio]{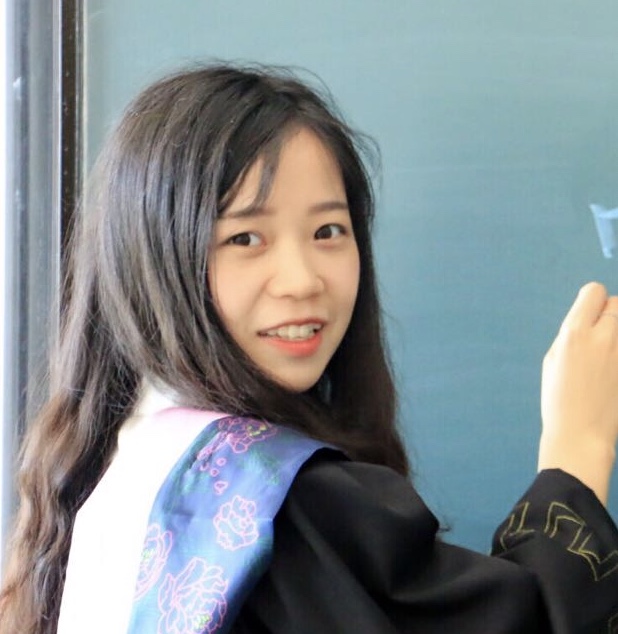}}]{Xueqi Yang} is a second year Ph.D. student in CS, North Carolina State University.  Her research interests include applying deep learning in software engineering and human-assisted AI algorithms. For more information, please visit \url{https://xueqiyang.github.io/}.
\end{IEEEbiography}\begin{IEEEbiography}[{\includegraphics[width=0.75in,clip,keepaspectratio]{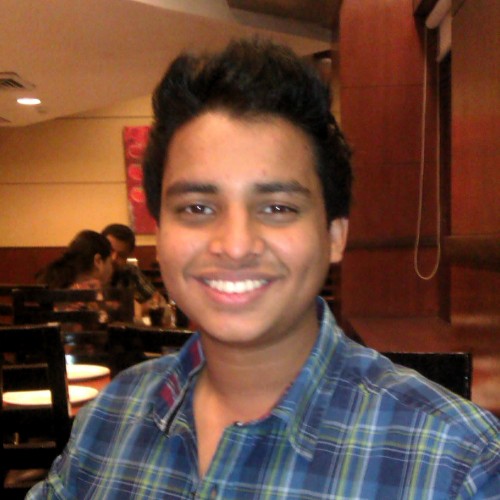}}]{Rishabh Agrawal}   is a second year master's student in CS, NC State University. He has three years of industry experience as a software engineer at Amazon. His research interests include machine learning for software engineering, data mining and deep learning. For more details, please visit http://tiny.cc/rishabhagrawal.
\end{IEEEbiography}\begin{IEEEbiography}[{\includegraphics[width=.75in,clip,keepaspectratio]{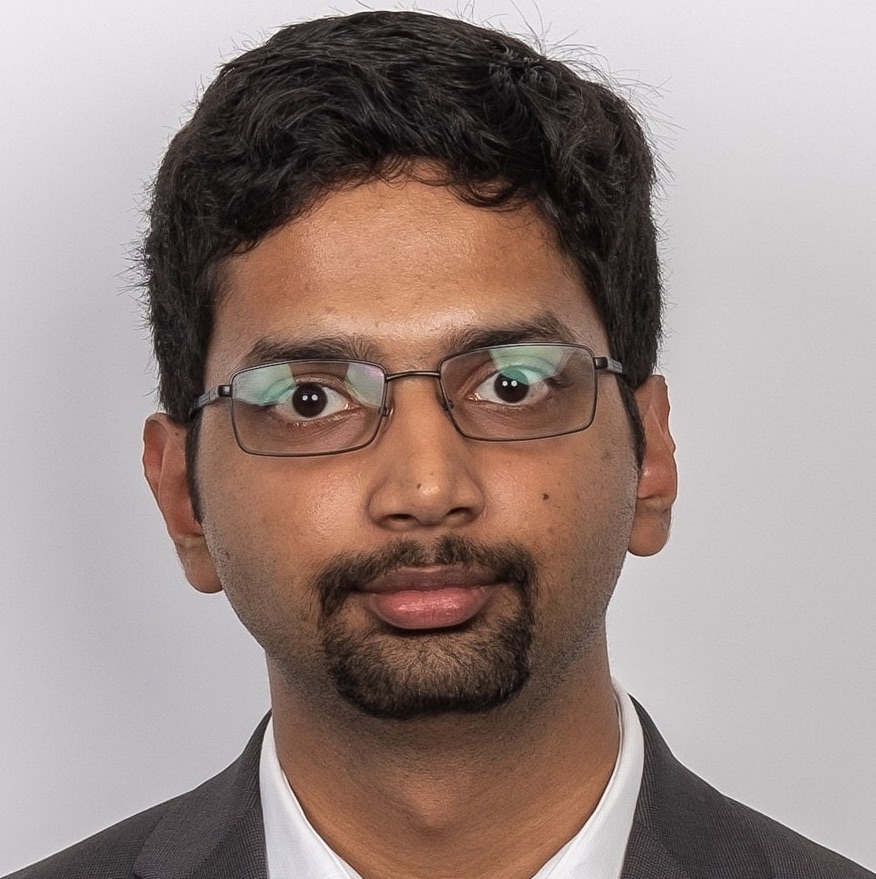}}]{Rahul Yedida} is a first-year PhD student in Computer Science at NC State University. His research interests include automated software testing and machine learning for software engineering. For more information, please visit \url{https://ryedida.me}.
\end{IEEEbiography}\begin{IEEEbiography}[{\includegraphics[width=0.75in,clip,keepaspectratio]{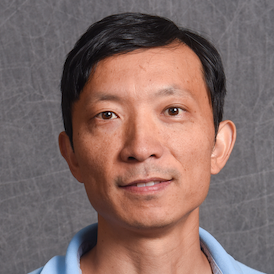}}]{Xipeng Shen} is a Professor in the CS,  NC State University.  He is an ACM Distinguished Member,  and a senior member of IEEE.   His primary research interest lies in the fields of Programming Systems and Machine Learning, emphasizing inter-disciplinary problems and approaches. 
\end{IEEEbiography}\begin{IEEEbiography}[{\includegraphics[width=0.75in,clip,keepaspectratio]{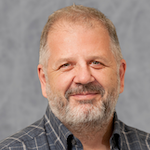}}]{Tim Menzies} (IEEE Fellow, Ph.D. UNSW, 1995)
is a Professor in CS at NC State University,  
where he teaches software engineering,
automated software engineering,
and programming languages.
His research interests include software engineering (SE), data mining, artificial intelligence, and search-based SE, open access science. 
For more information,  please visit \url{http://menzies.us}.
\end{IEEEbiography}

\end{document}